\documentclass[preprint,
nofootinbib,
longbibliography,
amsmath,amssymb,
aps,
prd
]{revtex4-2}

\usepackage[utf8]{inputenc}
\usepackage{amsmath}
\usepackage{amsfonts,dsfont}
\usepackage{mathrsfs}
\usepackage{cancel} 
\usepackage{amssymb,ulem}
\usepackage{amsmath}
\usepackage{romannum}
\usepackage{subfigure}
\usepackage{adjustbox}
\usepackage{dcolumn}
\usepackage{graphicx}
\usepackage{bm}
\usepackage{enumerate}
\usepackage[colorlinks=true,linkcolor=blue,urlcolor=blue,filecolor=black,
citecolor=red,pdfstartview=FitV,pdftitle={},pdfsubject={},pdfkeywords={},
pdfpagemode=None,bookmarksopen=true]{hyperref}

\usepackage[dvipsnames]{xcolor}
\usepackage{float}
\usepackage{enumitem}  
\usepackage{bbold}
\usepackage{tikz}
\usepackage{multirow}
\usepackage{longtable}
\usepackage{supertabular,rotating,adjustbox}
\begin{document}
	\preprint{APS/PRD}
	\title{\boldmath 
		B Anomalies  in  Two Higgs Doublet Model with Flavor Symmetry}
	\author{Qiaoyi Wen}
	\author{Fanrong Xu}
	\email{fanrongxu@jnu.edu.cn}
	
	\affiliation{Department of Physics, College of Physics $\&$ Optoelectronic Engineering,        
	Jinan University, Guangzhou 510632, P.R. China}

\begin{abstract} 

The long-standing flavor anomalies in $b \to s \ell^+ \ell^-$ and $b \to c \ell \nu$ persist. In this work, we explore a specific flavor-gauged two-Higgs doublet model (FG2HDM) extended by a scalar singlet, with an imposed $U(1)$ flavor symmetry. Compared to the Standard Model (SM), four additional scalars and one neutral gauge boson $Z'$ are added to the particle spectrum.
The special Yukawa coupling matrices, due to the $U(1)$ symmetry, lead to flavor-changing neutral Higgs (FCNH) interactions uniquely occurring in the down-type quark sector, while the new gauge boson $Z'$ induces flavor-changing neutral current (FCNC) interactions in the down-type quark sector as well. The two distinct types of anomalies can be accommodated simultaneously in such a model.
Incorporating scalar and vector contributions to $b \to s \ell^+ \ell^-$, $B_s\to \mu^+ \mu^-$, 
$B_s^0$-$\bar{B}_s^0$ mixing, charged Higgs contributions to $b \to c \ell \nu$, 
along with electroweak precision observables  (EWPOs) and Higgs fermionic decays, 
we investigate the parameter space in FG2HDM and find substantial room in the solution space. 
Specifically, in the SM-like model setup, the data suggests that $m_{Z'} < 450\, \text{GeV}$, $\tan\beta < 28$, and $g'/m_{Z'} > 3.5 \times 10^{-4} \, \text{GeV}^{-1}$, while there are fewer constraints on heavy scalars, which can be further tested by a combination of other observables.

\end{abstract}
\maketitle
\clearpage
\newpage
\pagenumbering{arabic}
\section{Introduction}
\label{sec: Intro}

Although the anomalies in $B$ meson decays have persisted for over a decade, serving as an indirect way to explore New Physics (NP) beyond the Standard Model (BSM) and attracting significant attention both experimentally and theoretically, the effort to precisely identify these anomalies continues.

The SM-like $R_{K^{(\ast)}}$ measured by LHCb at the end of 2022 \cite{LHCb:2022vje}, and later observed with large uncertainty by CMS as well \cite{CMS:2024syx}, indicated that the emergence of NP in terms of a violation of Lepton Flavor Universality (LFU) was not that promising. However, the possibility in binned branching fractions and polarization observables $P'_5$ still exists. LHCb performed a systematic $q^2$ {unbinned} amplitude analysis of $B^0 \to K^{\ast0} \mu^+ \mu^-$ at the end of 2023 \cite{LHCb:2023gel,LHCb:2023gpo} and found the deviation of $C_9$ from the SM to be about $1.9~\sigma$ and that the non-local operator is not sufficient to explain the deviation. In the latest {unbinned} amplitude analysis of May 2024 \cite{LHCb:2024onj,Hadavizadeh:2024yro}, using the Breit-Wigner improved method \cite{LHCb:2016due,Cornella:2020aoq}, LHCb obtained $C_9$ with a $2.1\sigma$ deviation from the SM prediction. The ongoing effort indicates that the NP opportunity in $b \to s \ell^+ \ell^-$ has not yet been concluded \cite{Carus:2024wra}.

Progress has also been made in the $b\to c\ell \bar{\nu}$ sector. Recently, LHCb reported that
		$R_{D^+}^{\rm LHCb} = 0.249 \pm 0.043 (\text{stat}) \pm 0.047 (\text{syst}), $ and 
		$R_{D^{\ast+}}^{\rm LHCb} = 0.402 \pm 0.081 (\text{stat}) \pm 0.085 (\text{syst})$
with a correlation coefficient $\rho = -0.39$ \cite{LHCb:2024jll}. The experimental uncertainty of $R_{D^\ast}$ remains considerable, while the error of $R_D$ decreased, pushing $R_D$ closer to the SM prediction than previously observed \cite{LHCb:2023zxo}. On the other hand, Belle II \cite{Belle-II:2024ami} announced their preliminary results of $R_{D^{(\ast)}}$ this year as
	$R_{D^{\ast}}^{\rm Belle} = 0.262^{+0.041}_{-0.039} (\text{stat})^{+0.035}_{-0.032} (\text{syst})$. 
\footnote{
The most recent measurements of $R_{D^{(\ast)}}$ reported by the Belle II in 2025, during the review process of this paper, yield a combined result that deviates from the SM prediction by  $1.7$ standard deviations, taking into account the correlation between the two observables~\cite{Belle-II:2025yjp}.
}
The preliminary world averages of $R_{D^{(\ast)}}$ for Moriond 2024, offered by HFLAV \cite{HFLAV:2022esi}, with the above Belle II result included \cite{Belle-II:2024ami}, are given as
\begin{equation}
	R_{D^{\ast}}^{\rm avg} = 0.287 \pm 0.012,~
	R_{D}^{\rm avg} = 0.342 \pm 0.026,
\end{equation}
with the same correlation of $-0.39$. Hence, the discrepancy between the SM prediction and the averaged value is at the level of $3.31\sigma$ \cite{HFLAV:2022esi}.  In addition, the longitudinal polarization fraction of the $D^\ast$ meson in $B^0\to D^{\ast-}\tau^+\nu_\tau$ provides complementary information to $R_{D^{\ast}}$ and has triggered experimental interest. However, the overall $q^2$ range integrated result reported by LHCb \cite{LHCb:2023ssl}, given as $F_L^{D^\ast} = 0.43 \pm 0.06 \pm 0.03$ by the end of 2023, was found to be SM compatible.
Recently, $J/\psi$ has also been employed as a supplement to $D^{(\ast)}$ mesons in the $b\to c\ell\nu$ process. 
A  preliminary combined result of $R_{J/\psi}$ announced by the CMS collaboration in August 2024  gives
$R_{J/\psi}^{\rm CMS} = 0.49\pm0.09 (\text{stat.})\pm0.25(\text{syst.})$, 
with large uncertainty that less information can be extracted at this moment \cite{Pacey:2024ppv,CMS:2024uyo}. 
Therefore, $R_{D^{\ast}}$ is still the main player of NP in $b\to c \ell {\nu}$.

It is commonly known that the two types of anomalies, occurring in $b \to s \ell^+ \ell^-$ and $b \to c \ell \nu$, are induced by neutral current and charged current interactions, respectively. This indicates that NP is encoded in distinct sectors and can be explored separately. In addition to model-independent global fits in $b \to s \ell^+ \ell^-$ \cite{Greljo:2022jac,Ciuchini:2022wbq,Wen:2023pfq,Altmannshofer:2023uci,Hurth:2023jwr,Bordone:2024hui,LHCb:2023gel,LHCb:2023gpo,LHCb:2024onj} and $b \to c \ell \nu$ \cite{Aebischer:2022oqe,Capdevila:2023yhq,DAlise:2024qmp,Iguro:2024hyk}, various specific NP models have also been discussed, including the $Z'$ models \cite{Buras:2013qja, Crivellin:2015era, Han:2019diw, Calibbi:2019lvs, Crivellin:2020oup, Ko:2021lpx, Nagao:2022osm, Allanach:2019iiy, Altmannshofer:2019xda, Crivellin:2022obd, Alok:2022pjb, Allanach:2023uxz, Athron:2023hmz, Loparco:2024olo} and leptoquark models \cite{Sahoo:2015wya, Becirevic:2016oho, Cline:2017aed, Allanach:2022iod, Belanger:2022kvj, Chen:2023wpb} to address neutral current anomalies, and 
{charged Higgs \cite{Crivellin:2012ye, Fajfer:2012jt, Celis:2012dk, Iguro:2017ysu, Iguro:2022uzz, Kumar:2022rcf, Iguro:2023jju}} and others such as $W'$ \cite{Bhattacharya:2014wla, Greljo:2015mma, Babu:2018vrl, Carena:2018cow} to address charged current anomalies. 
{Furthermore, efforts to accommodate both types of anomalies with
model-independent or model-dependent methods can also be found in references such as \cite{Greljo:2015mma, Buttazzo:2017ixm, Bhattacharya:2016mcc, Boucenna:2016qad, Kumar:2018kmr, Iguro:2018qzf, Crivellin:2023sig, Gherardi:2020qhc, Bhaskar:2022vgk, Chen:2022hle, Crivellin:2022mff,Kumar:2022rcf,Chen:2024jlj}}.

The Two-Higgs-Doublet Model (2HDM) is a natural extension of the SM and is commonly classified into several types regarding its Yukawa structures. To reduce redundancy in Yukawa coupling matrices, extra symmetry can be introduced. Taking 2HDM-II as an example, it is imposed with a symmetry for its first scalar doublet and right-handed down quarks ($\Phi_1 \to -\Phi_1$, $d_R^i \to d_R^i$), but faces difficulty in explaining the $R_{D^{\ast}}$ anomaly \cite{BaBar:2012obs}. There are still efforts in variant 2HDMs, such as 2HDM-III with Cheng-Sher ansatz \cite{Chen:2013qta,Cheng:1987rs}, gauged 2HDM (G2HDM) \cite{Huang:2015wts,Huang:2015rkj,Liu:2024nkl}, and BGL 2HDM \cite{Branco:1996bq} 
with special assignment of flavor symmetry  and some early discussion can be found in reference such as 
\cite{Berezhiani:1989fs}.
Here, we will focus on the BGL type models, with global flavor-symmetry-based Yukawa matrix textures introduced. Although there exists scalar-mediated FCNC, it can be suppressed by small off-diagonal entries of the CKM matrix \cite{Branco:1996bq}. Recently, the way to localize this BGL flavor symmetry has been proposed \cite{Celis:2015ara} and developed \cite{Ordell:2019zws,Ordell:2020yoq}, in which a new gauge boson corresponding to the $U(1)$ gauge group in charge of flavor symmetry and an extra scalar singlet are introduced in addition to the original two doublets in 2HDM. In this work, we explore how both anomalies can be accommodated within such a flavor-gauged 2HDM (FG2HDM).

This paper is organized as follows. In Sec. \ref{sec:frame}, we introduce the main structure of FG2HDM. In Sec. \ref{sec: pheo}, the model contributions to 
$b \to s \ell^+ \ell^-$, $b \to c \ell \nu$ and $B_s^0$-$\bar{B}_s^0$ mixing 
as well as $Z\to f \bar{f}$ and 
$h\to f\bar{f}$
are calculated. A further numerical analysis and a survey on model parameter space are presented in Sec. \ref{sec: num}. We summarize the entire work in Sec. \ref{sec:con}. Some details are provided in the Appendices, where App. \ref{app:Quan_nums} shows exact quantum numbers for a workable FG2HDM, and App. \ref{app:rotation_m} gives details of rotation matrices in the model. In App. \ref{app:pheo_detail}, some related phenomenology formulas are exhibited.

\section{Model setup}
\label{sec:frame}

\subsection{Scalar Potential}
\label{sub:scalar}
The scalar potential in the FG2HDM, involving two scalar doublets and one complex singlet, is given by
\begin{eqnarray}
    V(\Phi_1,\Phi_2,S) &=& m_{11}^2 \Phi_1^\dagger\Phi_1 + m_{22}^2 \Phi_2^\dagger\Phi_2
    + \frac{\lambda_1}{2} (\Phi_1^\dagger\Phi_1)^2 + \frac{\lambda_2}{2} (\Phi_2^\dagger\Phi_2)^2 \nonumber\\
    && + \lambda_3 (\Phi_1^\dagger\Phi_1)(\Phi_2^\dagger\Phi_2) + \lambda_4 (\Phi_1^\dagger\Phi_2)(\Phi_2^\dagger\Phi_1) \nonumber\\
    && + m_S^2 |S|^2 + \lambda_S |S|^4 + \kappa_1 |\Phi_1|^2 |S|^2 + \kappa_2 |\Phi_2|^2 |S|^2 \nonumber\\
    && + \kappa_3 (\Phi_1^\dagger\Phi_2 S^2 + \Phi_2^\dagger\Phi_1 (S^\ast)^2),
\end{eqnarray}
in which 
$\Phi_i=\left(\phi_i^+, (\rho_i+ i \eta_i +v_i)/\sqrt{2}\right)^T$
denotes the $SU(2)_L$ scalar doublet, and the singlet is parameterized as
$S=(v_S+s_0+i\chi_0)/\sqrt{2}$.
CP-violating phases in the vacuum expectation values (VEV) are neglected. Due to different charge assignments of these doublets under the gauged BGL symmetry, the terms $m_{12}$ and $\lambda_{5}$ vanish. 
The mass of the physical pseudoscalar arises from the newly introduced singlet after symmetry breaking. By utilizing stationary equations 
($\partial_i V_0\equiv \frac{\partial V_0}{\partial v_i}=0,~i=1,2,S$) around their vacua 
\begin{equation}\label{eq:stationary_cond}
    \begin{aligned}
        &\left(2 m_{11}^2 v_1+\kappa _1 v_1 v_S^2+\kappa _3 v_2 v_S^2+\lambda _1 v_1^3+\lambda _3 v_2^2 v_1+\lambda _4 v_2^2 v_1\right)=0,\\
        &\left(2 m_{22}^2 v_2+\kappa _2 v_2 v_S^2+\kappa _3 v_1 v_S^2+\lambda _2 v_2^3+\lambda _3 v_1^2 v_2+\lambda _4 v_1^2 v_2\right)=0,\\
        &\left(2 m_S^2 v_S+2 \lambda_S v_S^3+\kappa _1 v_1^2 v_S+\kappa _2 v_2^2 v_S+2 \kappa _3 v_1 v_2 v_S\right)=0,
    \end{aligned}
\end{equation}
we can derive the mass terms for the scalars
\begin{align}
    & \mathcal{L}_{\phi^\pm} = \frac12 (v_1 v_2\lambda_4+\kappa_3v_S^2)  \left(\begin{array}{cc} \phi_1^-, & \phi_2^-\end{array}\right)
    \left(\begin{array}{cc}
        \frac{v_2}{v_1} & -1 \\
        -1 & \frac{v_1}{v_2}
    \end{array}\right)
    \left(\begin{array}{c}
        \phi_1^+\\ \phi_2^+
    \end{array}\right),
 \\
    & \mathcal{L}_{\eta} = \frac14 \kappa_3v_S^2 \left(\begin{array}{ccc} \eta_1, & \eta_2, & \chi_0\end{array}\right)
    \left(\begin{array}{ccc}
        \frac{v_2}{v_1} & -1 &-\frac{2v_2}{v_S}\\
        -1 & \frac{v_1}{v_2} & \frac{2v_1}{v_S}\\
        -\frac{2v_2}{v_S} & \frac{2v_1}{v_S} & \frac{4v_1v_2}{v_S^2}
    \end{array}\right)
    \left(\begin{array}{c}
        \eta_1\\ \eta_2 \\ \chi_0
    \end{array}\right),
\nonumber\\
    &\mathcal{L}_{\rho} = -\frac12\left(\begin{array}{cc} \rho_1, \rho_2, s_0 \end{array}\right)
    \left(\begin{array}{ccc}
        \lambda_1 v_1^2-\frac{1}{2}\kappa_3\frac{v_2}{v_1}v_S^2 
        & \lambda_{34} v_1 v_2+\frac{1}{2}\kappa_3v_S^2
        & v_S(v_1\kappa_1+v_2\kappa_3)\\
        \lambda_{34} v_1 v_2+\frac{1}{2}\kappa_3v_S^2 
        & \lambda_2 v_2^2-\frac{1}{2}\kappa_3\frac{v_1}{v_2}v_S^2
        & v_S(v_2\kappa_2+v_1\kappa_3)\\
        v_S(v_1\kappa_1+v_2\kappa_3)
        & v_S(v_2\kappa_2+v_1\kappa_3)
        & 2\lambda_Sv_S^2
    \end{array}\right)
    \left(
    \begin{array}{c}
        \rho_1 \\ \rho_2 \\ s_0
    \end{array}\right), \nonumber
\end{align}
where $\lambda_{34}=\lambda_3+\lambda_4$. Assuming CP-even neutral scalar mass terms are block diagonal, i.e., $-\kappa_3=\kappa_1v_1/v_2=\kappa_2v_2/v_1 $ as adopted in \cite{Ko:2016lai}, we have
\begin{equation}\label{eq:CPeven_m}
    \mathcal{L}_{\rho} = -\frac12\left(\begin{array}{cc} \rho_1, \rho_2, s_0 \end{array}\right)
    \left(\begin{array}{ccc}
        \lambda_1 v_1^2-\frac{1}{2}\kappa_3\frac{v_2}{v_1}v_S^2 
        & \lambda_{34} v_1 v_2+\frac{1}{2}\kappa_3v_S^2
        & 0\\
        \lambda_{34} v_1 v_2+\frac{1}{2}\kappa_3v_S^2 
        & \lambda_2 v_2^2-\frac{1}{2}\kappa_3\frac{v_1}{v_2}v_S^2
        & 0\\
        0
        & 0
        & 2\lambda_Sv_S^2
    \end{array}\right)
    \left(
    \begin{array}{c}
        \rho_1 \\ \rho_2 \\ s_0
    \end{array}\right).
\end{equation}
giving the CP-even scalars masses
\begin{equation}
    m_{h,H}^2 = \frac12 \left[\left(m+n\right) \mp \sqrt{\left(m-n\right)^2 + 4l^2}\right], \quad
    m_{H_S}^2 = 2\lambda_Sv_S^2,
\end{equation}
where $m,~n,~l$ are defined as
\begin{equation}
    m = \lambda_1v_1^2 - \frac12\kappa_3\frac{v_2}{v_1}v_S^2, \quad
    n = \lambda_2v_2^2 - \frac12\kappa_3\frac{v_1}{v_2}v_S^2, \quad
    l = \lambda_{34}v_1v_2 + \frac12\kappa_3v_S^2.
\end{equation}
Likewise, the mass and gauge eigenstates of other scalar fields can be related  via rotation matrices
\begin{equation}
    \begin{aligned}
        &\begin{pmatrix}
            G^\pm\\
            H^\pm
        \end{pmatrix}=U_{1}\begin{pmatrix}
        \phi_1^\pm\\
        \phi_2^\pm
    \end{pmatrix},~
        \begin{pmatrix}
            G^{0}\\
            G^{\prime0}\\
            H_A
        \end{pmatrix}=U_{2}\begin{pmatrix}
            \eta_1\\
            \eta_2\\
            \chi_0
        \end{pmatrix},~
        \begin{pmatrix}
            H\\
            h\\
            H_S
        \end{pmatrix}=U_{3}\begin{pmatrix}
            \rho_1\\
            \rho_2\\
            s_0
        \end{pmatrix},
    \end{aligned}
\end{equation}
where $ U_{1} $, $ U_{2} $, and $ U_{3} $ matrices are detailed 
in Appendix \ref{app:rotation_m}. 
We present the squared masses of the CP-odd mode and charged Higgs
explicitly as
\begin{equation}
    \begin{aligned}
        m^2_{G^\pm} = m^2_{G^0} = m^2_{G^{\prime0}} = 0, \quad
        m_{H_A}^2 = -\frac12\kappa_3\left(4v_1v_2+\frac{v^2v_S^2}{v_1v_2}\right), \quad
        m^2_{H^\pm} = -{\frac12}(\lambda_4 v^2 + \frac{\kappa_3v^2v_S^2}{v_1v_2}).
    \end{aligned}
\end{equation}
The Goldstone theorem is preserved and hence the masses of Goldstone bosons 
naturally vanish as shown above.

\subsection{Gauge Interactions}
\label{subsec:gauge}

The kinetic terms for the scalar fields in the Lagrangian are given by
\begin{equation}\label{eq:scalar_kinetic}
    \mathcal{L}_{GS} = \left(D_\mu\Phi_1\right)^\dagger\left(D^\mu\Phi_1\right) + \left(D_\mu\Phi_2\right)^\dagger\left(D^\mu\Phi_2\right) + \left(D_\mu S\right)^\dagger\left(D^\mu S\right),
\end{equation}
where the gauge covariant derivatives to the scalar fields are defined as
\begin{equation}
    \begin{aligned}
        D_\mu S &= \left(\partial_\mu - i g' X_S \hat{Z}'_\mu\right)S, \\
        D_\mu\Phi_j &= \left(\partial_\mu - i g_1 Y_j B_\mu - i g' X_{\Phi_j} \hat{Z}'_\mu - ig_2 \frac{\vec{\tau}}{2} \cdot \vec{W}_\mu \right)\Phi_j,
    \end{aligned}
\end{equation}
with hypercharge under $U(1)_Y$ denoted as $Y_1 = Y_2 = \frac{1}{2}$. The quantum numbers $\mathcal{Q}_j$ of $\Phi_j$, encoded in $X_{\Phi_j}$,  under the $U(1)'$ group are detailed in Appendix~\ref{app:Quan_nums}.

\subsubsection{Gauge Boson Mass}
\label{subsubsec:gauge_mass}

After spontaneous symmetry breaking, the mass terms for the gauge bosons are expressed as
\begin{equation}
    \mathcal{L}_m^G = \left(\begin{array}{ccc}
        B & W^3 & \hat{Z}'
    \end{array}\right) \tilde{M} \left(\begin{array}{c}
        B \\ W^3 \\ \hat{Z}'
    \end{array}\right) = \left(\begin{array}{ccc}
        A & Z & Z'
    \end{array}\right) M_d \left(\begin{array}{c}
        A \\ Z \\ Z'
    \end{array}\right),
\end{equation}
where the $\tilde{M}$ matrix takes the form
\begin{equation}\label{eq:gauge_mass_nondiag}
    \tilde{M} = \frac{1}{8} (g_1^2 + g_2^2) v^2
    \left(\begin{array}{ccc}
        \sin^2\theta_W & -\sin\theta_W\cos\theta_W & 2\sin\theta_W\Delta \\
        -\sin\theta_W\cos\theta_W & \cos^2\theta_W & -2\cos\theta_W\Delta \\
        2\sin\theta_W\Delta & -2\cos\theta_W\Delta & 4\delta
    \end{array}\right).
\end{equation}
In particular, the notations $\Delta$ and $\delta$ are defined as
\begin{equation}
    \begin{aligned}
        \Delta &= \sin\xi \left(\mathcal{Q}_1 \cos\beta^2 + \mathcal{Q}_2 \sin\beta^2\right), \\
        \delta &= \sin^2\xi \left(\mathcal{Q}_1^2 \cos^2\beta + \mathcal{Q}_2^2 \sin^2\beta + \mathcal{R}_v^2 \mathcal{Q}_s^2\right),
    \end{aligned}
\end{equation}
with $\mathcal{R}^2_v = \frac{v_S^2}{v^2}$, $\sin\theta_W = \frac{g_1}{\sqrt{g_1^2 + g_2^2}}$, and $\sin\xi = \frac{g'}{\sqrt{g_1^2 + g_2^2}}$. The determinant of $\tilde{M}$ is zero, implying that one eigenvalue (corresponding to the photon) is zero. The rotation matrix $U$ connecting the mass eigenstates and gauge eigenstates is given by
\begin{equation}
\label{eq:gauge_rotation}
	\left(\begin{array}{c}
		A \\ Z \\ {Z}' \end{array}\right)
	= U  \left(\begin{array} {c}
		B\\ W^3 \\ \hat{Z}' \end{array}\right), \quad
	U\equiv U^T_{\hat{Z}\hat{Z}'}U^{T}_{A\hat{Z}}=\left(\begin{array}{ccc}
		\cos\theta_{W} & \sin\theta_{W} & 0 \\
		-\sin\theta_{W}\cos\theta_{2}^\prime & \cos\theta_{W} \cos\theta_{2}^\prime& \sin\theta_{2}^\prime\\
		\sin\theta_{W}\sin\theta_{2}^\prime& -\cos\theta_{W}\sin\theta_{2}^\prime & \cos\theta_{2}^\prime
	\end{array}\right),
\end{equation}
 with the matrices $U_{\hat{Z}\hat{Z}'}$ and $U_{A\hat{Z}}$ detailed in Appendix~\ref{app:rotation_m}. The diagonalized mass matrix $M_d$ is then given by
\begin{equation}
    M_d = \frac{1}{8} (g_1^2 + g_2^2) v^2
    \left(\begin{array}{ccc}
        0 & 0 & 0 \\
        0 & \mu^-_Z & 0 \\
        0 & 0 & \mu^+_{Z'}
    \end{array}\right),
\end{equation}
where $\mu_{Z^{(\prime)}}^{\mp} = \frac12 \left[1 + 4\delta \mp \sqrt{(4\delta - 1)^2 + 16\Delta^2}\right]$. 
Here
we have assumed $\mathcal{R}_v$ is sufficiently large, making the $Z'$ boson heavier than the SM $Z$ boson.

\subsubsection{Fermion Kinetic Terms}
\label{subsubsec:fermion_kinetic}

The interactions of the new gauge bosons with fermions are obtained from
\begin{equation}
    \begin{aligned}
        \mathcal{L}_\text{FG} &= \overline{L_{L}^0} i\gamma^{\mu} D_\mu L_{L}^0 + \overline{Q_{L}^0} i\gamma^\mu D_\mu Q_{L}^0 + \overline{e_{R}^0} i\gamma^\mu \left(\partial_\mu - ig_1 Y B_\mu - ig' X_{\ell R} \hat{Z}_\mu\right) e_{R}^0 \\
        &+ \overline{u_{R}^0} i\gamma^{\mu} \left(\partial_\mu - ig_1 Y B_\mu - ig' X_{uR} \hat{Z}_\mu\right) u_{R}^0 + \overline{d_{R}^0} i\gamma^\mu \left(\partial_\mu - ig_1 Y B_\mu - ig' X_{dR} \hat{Z}_\mu\right) d_{R}^0,
    \end{aligned}
\end{equation}
where fermion fields with superscript $0$ denote fields in the gauge eigenstate. The neutral current terms in the mass basis are then given by
\begin{equation}\label{eq:Zff}
    \begin{aligned}
        \mathcal{L}_\text{FG} = & \quad
         e Q_f A_\mu \bar{f} \gamma^\mu f \\
        &+ \frac{g_2 \cos\theta_{2}^\prime}{\cos\theta_{W}} Z_\mu \bar{f} \left[(I_3^f - Q_f \sin^2\theta_{W}) \gamma^\mu P_L - Q_f \sin^2\theta_{W} \gamma^\mu P_R\right] f \\
        &- \frac{g_2 \sin\theta_{2}^\prime}{\cos\theta_{W}} Z_\mu' \bar{f} \left[(I_3^f - Q_f \sin^2\theta_{W}) \gamma^\mu P_L - Q_f \sin^2\theta_{W} \gamma^\mu P_R\right] f \\
        &+ g' \sin\theta_{2}^\prime Z_\mu \bar{f} \left[\mathcal{Q}_{fL} \gamma^\mu P_L + \mathcal{Q}_{fR} \gamma^\mu P_R\right] f \\
        &+ g' \cos\theta_{2}^\prime Z_\mu' \bar{f} \left[\mathcal{Q}_{fL} \gamma^\mu P_L + \mathcal{Q}_{fR} \gamma^\mu P_R\right] f,
    \end{aligned}
\end{equation}
where $f = u, d, \nu, \ell$, $I_3^f$ is the weak isospin third component, and $Q_f$ is the electric charge. The couplings to fermions of different chiralities are determined by the $U(1)'$ charges, given generally as
\begin{equation}
    \begin{aligned}
        \mathcal{Q}_{fL} &= U^\dagger_{fL} X_{fL} U_{fL}, \\
        \mathcal{Q}_{fR} &= U^\dagger_{fR} X_{fR} U_{fR},
    \end{aligned}
\end{equation}
where $X_{fR}$ structures are discussed in Appendix~\ref{app:Quan_nums}, and $U_{fL(R)}$ are rotation matrices connecting fermions in the mass eigenstate and weak eigenstate via
\begin{equation}
    f^0_{L(R)} = U_{fL(R)} f_{L(R)}.
\end{equation}
Hence the CKM and PMNS matrices are defined as
\begin{equation}
    V \equiv V_{\rm CKM} = U_{uL}^\dagger U_{dL}, \quad
    U \equiv V_{\rm PMNS} = U_{\nu L}^\dagger U_{\ell L},
\end{equation}
which play crucial roles in gauge interactions and fermion-scalar interactions.

\subsection{Yukawa interaction}
\label{sub: yukawa}
The Yukawa interactions in FG2HDM, including both quark and lepton sectors,  are generally 
in the form of
\begin{eqnarray}
    -\mathcal{L}_Y &=& \overline{Q^0_L}(Y_1^d\Phi_1 + Y_2^d\Phi_2)d_R^0 + \overline{Q^0_L}(Y_1^u\tilde{\Phi}_1 + Y_2^u\tilde{\Phi}_2)u_R^0 \nonumber \\
    &&+ \overline{L^0_L}(Y_1^\ell{\Phi_1} + Y_2^\ell{\Phi_2})e_R^0 + \overline{L^0_L}(Y_1^\nu\tilde{\Phi}_1 + Y_2^\nu\tilde{\Phi}_2)\nu_R^0 + \text{h.c.},
\end{eqnarray}
where the superscripts denote the fermion fields in the aforementioned gauge eigenstate.
Generically there exist eight $3\times 3$ Yukawa matrices in total, including the two
related to neutrino which provide Dirac neutrino mass.

\subsubsection{Fermion Mass and Yukawa Matrices}
After spontaneous symmetry breaking, fermion mass terms take the form
\begin{equation}
    -\mathcal{L}_m = \bar{u}_L M_u u_R + \bar{d}_L M_d d_R + \bar{\ell}_L M_\ell \ell_R + \bar{\nu}_L M_\nu \nu_R + \text{h.c.},
\end{equation}
where $M_f$ ($f=u,d,\ell,\nu$) are diagonal mass matrices, explicitly giving
$M_u = \textrm{diag}(m_u, m_c, m_t)$,
$M_d = \textrm{diag}(m_d, m_s, m_b)$,
$M_\ell = \textrm{diag}(m_e, m_\mu, m_\tau)$,
$M_\nu = \textrm{diag}(m_{\nu_1}, m_{\nu_2}, m_{\nu_3})$.
Note the physical fermion mass is rotated from generally non-diagonal $\tilde{M}_f$ as
\begin{align}
    M_f &= U_{fL}^\dagger \tilde{M}_f U_{fR}, \\
    \tilde{M}_f &= \frac{1}{\sqrt{2}}(v_1 Y_1^f + v_2 Y_2^f).
\end{align}

It is convenient to introduce the auxiliary matrices defined as
\begin{equation}
	N_f=\frac{1}{\sqrt{2}}U_{fL}^\dagger(v_1 Y_2^f - v_2 Y_1^f) U_{fR}.
\end{equation}
for the description of  flavor changing  processes involving scalars. 
Since two types of Yukawa matrices $ Y_1 $, $ Y_2 $ in FG2HDM 
are involved,
 the form of $N_f$ is hence determined by the formats of $Y_i^f$, which further can be regarded as a consequence of flavor symmetry in this work. 
Without loss of generality,
the $N_f$ matrices can be connected to mass matrices via
\begin{equation}
	N_f= -\frac{v_2}{v_1} M_f + \frac{v_2}{\sqrt{2}} \left(\frac{v_2}{v_1} + \frac{v_1}{v_2} \right)
	U_{fL}^\dagger
	Y_2^f U_{fR}.
\end{equation}
Note the Yukawa coupling matrices have not yet received restrictions so far, and hence
 keep their most general structure.
 We will show in the following, 
the desired  FCNC or FCCC  controlled by CKM matrix can be achieved by  special texture of Yukawa coupling matrices, which is a natural consequence of 
special flavor symmetries.

\subsubsection{The specific texture for Yukawa coupling matrices}

Assigned specific $U(1)^\prime$ quantum numbers detailed in Appendix \ref{app:Quan_nums}, we obtain the following texture for quark Yukawa matrices,
\begin{equation}
    Y_1^u = \left( \begin{array}{ccc}
        * & * & 0 \\
        * & * & 0 \\
        0 & 0 & 0 
    \end{array} \right), \quad
    Y_2^u = \left( \begin{array}{ccc}
        0 & 0 & 0 \\
        0 & 0 & 0 \\
        0 & 0 & * 
    \end{array} \right), \quad
    Y_1^d = \left( \begin{array}{ccc}
        * & * & * \\
        * & * & * \\
        0 & 0 & 0 
    \end{array} \right), \quad
    Y_2^d = \left( \begin{array}{ccc}
        0 & 0 & 0 \\
        0 & 0 & 0 \\
        * & * & * 
    \end{array} \right), \label{eq:quarkYukawa}
\end{equation}
where '$*$' denotes a non-zero arbitrary number. This structure of Yukawa matrices leads to the form of $N_f$ as:
\begin{align}
    N_u &= -\frac{v_2}{v_1} \,{\rm diag}(m_u, m_c, 0) + \frac{v_1}{v_2} \, {\rm diag}(0, 0, m_t), \nonumber \\
    (N_d)_{ij} &= -\frac{v_2}{v_1} (M_d)_{ij} + \left(\frac{v_2}{v_1}+\frac{v_1}{v_2}\right) (V^\dagger_{\rm CKM})_{i3} (V_{\rm CKM})_{3j} (M_d)_{jj}, \label{eq:Nq}
\end{align}
where the dummy index is not summed. Clearly, tree-level FCNC interaction induced by scalar does not occur in the up-type quark sector.

Likewise, by imposing the following lepton Yukawa textures,
\begin{equation}
    Y_1^\ell = \left( \begin{array}{ccc}
        0 & 0 & 0 \\
        0 & * & 0 \\
        0 & 0 & * 
    \end{array} \right), \quad
    Y_2^\ell = \left( \begin{array}{ccc}
        * & 0 & 0 \\
        0 & 0 & 0 \\
        0 & 0 & 0 
    \end{array} \right), \quad
    Y_1^\nu = \left( \begin{array}{ccc}
        0 & 0 & 0 \\
        0 & 0 & 0 \\
        * & * & * 
    \end{array} \right), \quad
    Y_2^\nu = 0, \label{eq:lepYukawa}
\end{equation}
we find the coupling matrices among leptons and scalars to be diagonal, giving
\begin{align}
    N_\nu &= -\frac{v_2}{v_1} M_\nu, \nonumber \\
    N_\ell &= -\frac{v_2}{v_1} \textrm{diag}(0, m_\mu, m_\tau) + \frac{v_1}{v_2} \textrm{diag}(m_e, 0, 0). \label{eq:Nl}
\end{align}
Apparently,  
$N_d$ uniquely introduces tree-level flavor-changing scalar-mediated interactions for down-type quark
among all the interactions between fermions and scalar.

\subsubsection{The interactions between fermions and scalars}

We have learned from Sec. \ref{sub:scalar} that after spontaneous symmetry breaking (SSB), five 
physical scalar particles ($h$, $H$, $H_S$, $H_A$, $H^\pm$) remain in the FG2HDM. In the mass basis, the interaction among fermions and physical scalars can be written as
\begin{equation}\label{eq:Yukawa}
	\begin{aligned}
		-\mathcal{L} &= \frac{\sqrt{2}}{v} H^+ \left[ \bar{u}\left(V_{\rm CKM} N_d P_R - N_u^\dagger V_{\rm CKM} P_L\right)d 
		+ \bar{\nu}\left(V_{\rm PMNS} N_\ell P_R - N_\nu^\dagger V_{\rm PMNS} P_L\right)\ell
		\right]+\text{h.c.} \\
		&\quad+\frac{1}{v}\left[\sin(\alpha-\beta)M_f-\cos(\alpha-\beta)N_f\right]h\bar{f}f
		-\frac{1}{v}\left[\cos(\alpha-\beta)M_f+\sin(\alpha-\beta)N_f\right]H\bar{f}f \\
		&\quad+\frac{i}{v}\left[U_\gamma^{21} N_d\right]H_A\bar{d}\gamma_5d
		-\frac{i}{v}\left[U_\gamma^{21} N_u\right]H_A\bar{u}\gamma_5u
		+\frac{i}{v}\left[U_\gamma^{21} N_\ell\right]H_A\bar{\ell}\gamma_5\ell,
	\end{aligned}
\end{equation}
The CP-even scalar block mass matrix in Eq. \eqref{eq:CPeven_m} determines the absence of $H_S$, while $H_A$ enters through rotation. These interactions lay the groundwork for further phenomenological discussions.

\section{Phenomenology}
\label{sec: pheo}

The semi-leptonic processes within the $B$ meson system provide crucial windows for investigating NP. 
The multiple scalars emerged in FG2HDM take part in various interactions at high energy. For instance, 
the charged Higgs plays a role in $b \to c \ell \nu$ while neutral scalar may contribute to $b \to s \ell \ell$ process.
In this section, we will explore in detail NP contribution to semileptonic $B$ decays as well as $B_s^0-\overline{B}^0_s$ mixing.

\subsection{\texorpdfstring{$ b\to c\ell\nu $}{}}
By integrating out the heavy scalar, the quark-level decay \( b \to c \bar{\ell} \nu \) can be depicted by the following effective Hamiltonian \cite{Murgui:2019czp,Iguro:2024hyk}:\footnote{
The adopted Hamiltonian form is for the convenience of extensions to incorporate interactions induced by
right-handed neutrinos \cite{Robinson:2018gza,Greljo:2018ogz}.}
\begin{equation}
    \mathcal{H}= 2\sqrt{2}G_FV_{cb}\left[(1+C_{V_L})\mathcal{O}_{V_L}
    +C_{V_R}\mathcal{O}_{V_R}+C_{S_L}\mathcal{O}_{S_L}+C_{S_R}\mathcal{O}_{S_R}+C_{T}\mathcal{O}_{T}
    \right]+\text{h.c.}
\end{equation}
with four effective operators given by
\begin{align}
    &\mathcal{O}_{V_{L,R}} = (\bar{c}\gamma^\mu P_{L,R} b)(\bar{\ell} \gamma_\mu P_L \nu), \quad
    \mathcal{O}_{S_L} = (\bar{c} P_L b)(\bar{\ell} P_L \nu),\nonumber\\
    &\mathcal{O}_{S_R} = (\bar{c} P_R b)(\bar{\ell} P_L \nu), \quad
    \mathcal{O}_{T} = (\bar{c} \sigma^{\mu\nu} P_L b)(\bar{\ell} \sigma_{\mu\nu} P_L \nu).
\end{align}
Note all coefficients here represent  pure NP contributions since contributions from the Standard Model (SM)
 have already been factorized out. In FG2HDM the contribution from exotic charged Higgs \( H^+ \)  is naturally generated at tree level, 
 while the other contributions from new \( Z' \) and neutral \( H \)  emerge from loop level and have
 been neglected. Hence only scalar operators can be generated here, and  Wilson coefficients at  EW scale are explicitly given as
\begin{equation}
\label{eq:C_SLR}
	\footnotesize
    \begin{aligned}
        C_{S_L}^{\tau3}&=-\frac{2}{v^2 m_{H^\pm}^2}\left(N_u^\dagger V_{\text{CKM}}\right)_{23}(N_\ell)^\dagger_{\tau3}\Big/(-2\sqrt{2}G_FV_{cb})
        =\frac{1}{m_{H^\pm}^2}\left(-\frac{v_2}{v_1}m_c\right)\left(-\frac{v_2}{v_1}m_\tau\right)
        =\frac{ m_c m_\tau}{m_{H^\pm}^2}\tan^2\beta,\\
        C_{S_R}^{\tau3}&=\frac{2}{v^2 m_{H^\pm}^2}\left(V_{\text{CKM}}N_d\right)_{23}(N_\ell)^\dagger_{\tau3}\Big/(-2\sqrt{2}G_FV_{cb})
        =-\frac{m_b m_\tau}{m_{H^\pm}^2}\tan^2\beta .
    \end{aligned}
\end{equation}
 By retaining $C_{S_R}^{\tau 3}$, an additional contribution via $C_{S_L}^{\tau 3}$ enters compared to the early estimation of 2HDM-II \cite{Hou:1992sy}. 
 For the comparison with the complete  2HDM-II \cite{Atkinson:2021eox} result, a significant 
 difference exists in the relative sign 
 between $C_{S_R}^{\tau 3}$ and  $C_{S_L}^{\tau 3}$.
 This alleviates the tension between the 2HDM (especially 2HDM-II) and the $R_{D^{(*)}}$ measurement \cite{BaBar:2012nus}.
In order to calculate low energy observables,  
QCD renormalization group (RG) evolution of \( C_{S_{L,R}} \) in the leading approximation 
is considered.   
Running from \( \Lambda=100 \) GeV to \( \mu=m_b \), we have the Wilson coefficients at low energy 
\begin{equation}\label{eq:Cs_running}
    \begin{aligned}
        C_{S_{L,R}}(\mu)&=\underbrace{\left[\frac{\alpha_s(m_t)}{\alpha_s(m_b)}\right]^{-8/(2\beta_0^{(5)})}}_{\lambda_c}C_{S_{L,R}}(\Lambda),
    \end{aligned}
\end{equation}
with the one-loop \( \beta \)-function \( \beta_0^{(n)}= 11-2n/3 \), \( \lambda_c\approx 1.41 \) as adopted in \cite{Dorsner:2013tla,Freytsis:2015qca}.

Observables related to \( b \to c \ell \nu \) process, 
associated with  Wilson coefficients evaluated at the \( \mu=m_b \),
have the following numerical forms  \cite{Duan:2024ayo},
\begin{equation}
	\begin{aligned}
		        & R_D = R_D^{\rm SM} \left[
		        1 + 1.07|C_{S_L}^{\tau3} + C_{S_R}^{\tau3}|^2 + 1.528\Re[C_{S_L}^{\tau3\ast} + C_{S_R}^{\tau3\ast}]\right],
		        \\
		        & R_{D^{*}} = R_{D^{*}} ^{\rm SM} \left[
		        1 + 0.043|C_{S_L}^{\tau3} - C_{S_R}^{\tau3}|^2 - 0.113\Re[C_{S_L}^{\tau3\ast} - C_{S_R}^{\tau3\ast}]
		        \right],
		        \\
		        & F_{L}^{D^\ast} =F_{L,\text{SM}}^{D^\ast}  \left(\frac{R_{D^\ast}}{R_{D^\ast}^\text{SM}}\right)^{-1}
		        \times \left(1 + 0.099|C_{S_L}^{\tau3}-C_{S_R}^{\tau3}|^2  -  0.263\Re[C_{S_L}^{\tau3\ast} - C_{S_R}^{\tau3\ast}]
		        \right),
		        \\
		        & P_{\tau}^{D^\ast} = P_{\tau,\text{SM}}^{D^\ast} \left(\frac{R_{D^\ast}}{R_{D^\ast}^\text{SM}}\right)^{-1}
		        \times \left( 1 - 0.082|C_{S_R}^{\tau3} - C_{S_L}^{\tau3}|^2 + 0.218\Re[C_{S_L}^{\tau3\ast} - C_{S_R}^{\tau3\ast}]
		        \right),
		        \\
		        & R_{J/\psi} =R_{J/\psi}^{\text{SM}}\times\left(
		        1 + 0.043|C_{S_L}^{\tau3} - C_{S_R}^{\tau3}|^2 - 0.113\Re\left[C_{S_L}^{\tau3\ast}-C_{S_R}^{\tau3\ast}\right] 
		        \right),
		        \\
		        & R_{\Lambda_c}=R_{\Lambda_c}^{\text{SM}}\times\left(
		        1 + 0.321\left(|C_{S_L}^{\tau3}|^2 + |C_{S_R}^{\tau3}|^2\right)
		        	+ 0.334\Re[C_{S_L}^{\tau3\ast}] 
		        	+ 0.497 \Re[C_{S_R}^{\tau3\ast}] 
		        	+ 0.513\Re[C_{S_L}^{\tau3}C_{S_R}^{\tau3\ast}]
		        \right),\\
		        & \mathcal{B}(B_c\to\tau\bar{\nu}) =
		        \mathcal{B}(B_c\to\tau\bar{\nu})_{\rm SM}
		        \times\left|1-4.35(C_{S_L}^{\tau3} - C_{S_R}^{\tau3})\right|^2,\\
		        & \mathcal{B}(B_u\to\tau\bar{\nu})=
		        9.30\left|1+3.85*(C_{S_R}^{\tau3}-C_{S_L}^{\tau3})|_{ub}\right|^2\times10^{-5},
	\end{aligned}
    \label{eq:RD}
\end{equation}
where $(C_{S_R}^{\tau3} - C_{S_L}^{\tau3})|_{ub}$ in the last equation indicates that the quark flavor in Eq.~\eqref{eq:C_SLR} should be adjusted accordingly, with $u$ assigned to $C_{S_L}^{\tau3}$ and $b$ to $C_{S_R}^{\tau3}$. 
Although the $b \to u\ell\nu$ process slightly deviates from the main focus of this subsection, we discuss it later due to its relevance to $H^\pm$. 

Recent advances in lattice QCD calculations of the $B \to D^{(\ast)}$ form factors~\cite{FermilabLattice:2021cdg,Aoki:2023qpa}
have led to updated estimates of 
semileptonic decays of  $B \to D^{(\ast)}$ ~\cite{Duan:2024ayo, Ray:2023xjn}.  
These developments refine earlier version of numerical formulas quoted in ~\cite{Iguro:2024hyk, Crivellin:2012ye}. In this work, we adopt the results presented in~\cite{Duan:2024ayo}
 since  the full $q^2$ range, combining the latest calculations from both lattice QCD in low recoil region and light-cone sum rules in large hadronic recoil~\cite{Cui:2023jiw}, has been incorporated.
 \footnote{
Some of the relevant numbers in Eq. \eqref{eq:RD}, which are not yet publicly available in the current version of the paper, have been provided through private communications with the authors and will appear in their forthcoming update.
}
For the leptonic decay $B_c\to \tau\bar{\nu}$, we still cite the formula adopted in ~\cite{Iguro:2024hyk}.
The other theoretical predictions and experimental measurements have been summarized and can be found in Sec.~\ref{sec: num}.

\subsection{\texorpdfstring{$ b\to s\ell\ell $}{}}

The effective Hamiltonian to describe $ b\to s \ell^+ \ell^-$ are given as
\begin{equation}
	\mathscr{H}=-\frac{4G_F}{\sqrt{2}}V_{tb}V^*_{ts}\frac{e^2}{16\pi^2}\sum_i\left(C_i \mathcal{O}_i +C'_i\mathcal{O}'_i\right)+\text{h.c.,}
\end{equation}
where the effective operators are defined as
{
\begin{equation}
	\begin{aligned}
		& \mathcal{O}_7=\frac{m_b}{e}(\bar{s}\sigma_{\mu\nu}P_R b)F^{\mu\nu},
		&
		&\mathcal{O}'_7=\frac{m_b}{e}(\bar{s}\sigma_{\mu\nu}P_L b)F^{\mu\nu},
		\\
		& \mathcal{O}_8=\frac{g_sm_b}{e^2}(\bar{s}\sigma_{\mu\nu}T^aP_R b)G^{\mu\nu}_a,
		&
		&\mathcal{O}'_8=\frac{g_sm_b}{e^2}(\bar{s}\sigma_{\mu\nu}T^aP_L b)G^{\mu\nu}_a,
		\\
		& \mathcal{O}_9=(\bar{s}\gamma_\mu P_L b)(\bar{\ell}\gamma^\mu \ell),
		&
		&\mathcal{O}'_9=(\bar{s}\gamma_\mu P_R b)(\bar{\ell}\gamma^\mu \ell),
		\\
		& \mathcal{O}_{10}=(\bar{s}\gamma_\mu P_L b)(\bar{\ell}\gamma^\mu\gamma_5 \ell),
		&
		&\mathcal{O}'_{10}=(\bar{s}\gamma_\mu P_R b)(\bar{\ell}\gamma^\mu \gamma_5\ell),
		\\
		& \mathcal{O}_{S}=m_b(\bar{s} P_R b)(\bar{\ell}  \ell),
		&
		&\mathcal{O}'_{S}=m_b(\bar{s} P_L b)(\bar{\ell}  \ell),
		\\
		& \mathcal{O}_{P}=m_b(\bar{s} P_R b)(\bar{\ell} \gamma_5 \ell),
		&
		&\mathcal{O}'_{P}=m_b(\bar{s} P_L b)(\bar{\ell} \gamma_5 \ell).
	\end{aligned}
\end{equation}
}
It is known that the decay branching fraction only receives contribution from $\mathcal{O}_{7,8,9,10}$ in SM while
more operators may contribute in NP models. In FG2HDM, 
Wilson coefficients of $ \mathcal{O}_{9,10}$ receives modifications due to the FCNC interaction induced by neutral
vector bosons, giving 
\begin{equation}\label{eq:C9C10}
	\begin{aligned}
		\Delta C_{9}&=-\frac{1}{N}\left[
		\frac{1}{m_{Z'}^2}\mathcal{A}_L^{bsZ'}\mathcal{B}^{Z'\ell\ell}
		+\frac{1}{m_{Z}^2}\mathcal{A}_L^{bsZ}\mathcal{B}^{Z\ell\ell}
		\right],\\
		\Delta C_{10}&=-\frac{1}{N}\left[
		\frac{1}{m_{Z'}^2}\mathcal{A}_L^{bsZ'}\mathcal{B}_5^{Z'\ell\ell}
		+\frac{1}{m_{Z}^2}\mathcal{A}_L^{bsZ}\mathcal{B}_5^{Z\ell\ell}
		\right],\\
	\end{aligned}
\end{equation}
with $ N= 2\sqrt{2}G_FV_{tb}V_{ts}^\ast e^2/(16\pi^2)$. The parameters 
$ \mathcal{A}_{L}^{bsZ^{(\prime)}} $, $ \mathcal{B}^{Z^{(\prime)}\ell\ell} $, and $ \mathcal{B}_{5}^{Z^{(\prime)}\ell\ell} $,
summarized in Appendix \ref{app:pheo_detail},
 represent the corresponding vertices mediated by $ Z/Z' $ with $ P_L$ in quark side, and $\gamma^\mu, $ and $ \gamma^\mu\gamma_5 $ in lepton side, respectively. The absence of $ \Delta C_{9}^{\prime}$ and $\Delta C_{10}^\prime$ can be interpreted 
as a consequence of diagonal $ \mathcal{Q}_{dR} $.

In addition to a modification of $\mathcal{O}_{9,10}$ contribution, scalar and pseudoscalar-type operators can be generated at leading-order.
Due to the interaction in Eq. \eqref{eq:Yukawa}, the FCNH information 
is encoded in Wilson coefficients of the forms,
\begin{equation}\label{eq:CSCp}
	\begin{aligned}
		\Delta C_{S}^\ell&=\Delta C_{S}^{\prime\ell}\\
		&=\frac{1}{N}\frac{2\sqrt{2}G_FV_{tb}V_{ts}^\ast}{\sin2\beta}
		\left\{\left(\frac{1}{m_H^2}-\frac{1}{m_h^2}\right)\left[\sin^2(\alpha-\beta)N_\ell+\sin(\alpha-\beta)\cos(\alpha-\beta)M_\ell\right]\right.
		\\
		&\hspace{4.0cm}\left.+\frac{1}{m_h^2}N_\ell\right\},
		\\
		\Delta C_{P}^\ell&=-\Delta C_{P}^{\prime\ell}
		=-\frac{1}{N}\frac{2\sqrt{2}G_FV_{tb}V_{ts}^\ast}{m_{H_A}^2\sin2\beta}\left[U_\gamma^{21}\right]^2 N_\ell \approx -\frac{1}{N}\frac{2\sqrt{2}G_FV_{tb}V_{ts}^\ast}{m_{H_A}^2\sin2\beta} N_\ell.
	\end{aligned}
\end{equation}

The Wilson coefficients of scalar and pseudoscalar operators should have the same anomalous dimension as quark masses
as stated in \cite{Crivellin:2013wna}, indicating that 
the  RGE Eq. \eqref{eq:Cs_running} is also applicable here.
In principle, the RGE correction of $ C_{9,10} $ running from high energy to $ m_b $ scale should be taken into account.
However,  such a correction is not sizable from the work in \cite{Hu:2016gpe} and hence is neglected
in the following numerical analysis.

\subsection{\texorpdfstring{$B_s-\bar{B}_s$}{} Mixing}

The mixing in $B_s^0$ system provides complementary information to semileptonic decay $b\to s \ell^+ \ell^-$ in view of pure quark transition $b\to s$. 
Constraints on $bs(Z'/H)$ vertices in  FG2HDM are aimed to be extracted from experimental data.

Within the state-of-the-art effective Hamiltonian approach, the Hamiltonian in BMU basis \cite{Buras:2000if,Buras:2001ra} is given by
\begin{equation}
	\mathcal{H}_\text{eff}^{B\bar{B}} = N_{B\bar{B}} \sum_{i} C_i(\mu) \mathcal{O}_i,
\end{equation}
where the four-quark operators are defined as
\begin{equation}
	\begin{aligned}
		\mathcal{O}_\text{VLL} &= \left(\bar{b} \gamma_\mu P_L q \right) \left( \bar{b} \gamma^\mu P_L q \right), & \quad
		\mathcal{O}_\text{VRR} &= \left( \bar{b} \gamma_\mu P_R q \right) \left( \bar{b} \gamma^\mu P_R q \right), \\
		\mathcal{O}_\text{LR}^1 &= \left( \bar{b} \gamma_\mu P_L q \right) \left( \bar{b} \gamma^\mu P_R q \right), & \quad
		\mathcal{O}_\text{LR}^2 &= \left( \bar{b} P_L q \right) \left( \bar{b} P_R q \right), \\
		\mathcal{O}_\text{SLL}^1 &= \left( \bar{b} P_L q \right) \left( \bar{b} P_L q \right), & \quad
		\mathcal{O}_\text{SRR}^1 &= \left( \bar{b} P_R q \right) \left( \bar{b} P_R q \right), \\
		\mathcal{O}_\text{SLL}^2 &= -\left( \bar{b} \sigma_{\mu\nu} P_L q \right) \left( \bar{b} \sigma^{\mu\nu} P_L q \right), & \quad
		\mathcal{O}_\text{SRR}^2 &= -\left( \bar{b} \sigma_{\mu\nu} P_R q \right) \left( \bar{b} \sigma^{\mu\nu} P_R q \right),
	\end{aligned}
\end{equation}
and the constant is $N_{B\bar{B}} \equiv \frac{1}{16 \pi^2}G_F^2 M_W^2 (V_{tb}^\ast V_{ts})^2$.

In FG2HDM, new contributed Wilson coefficients to calculate the mass split 
$\Delta M_s=2|M_{12}| \equiv 2|\langle B^0| \mathcal{H}_{\rm eff}^{B\bar{B}} | \bar{B}^0 \rangle|$,
generated by integrating out heavy particles, are calculated to be
\begin{equation}
\label{eq:del_F=2}
	\begin{aligned}
		\Delta C_\text{VLL} &= \frac{g^{\prime 2}}{4} \left( \frac{\sin^2 \theta_{2}'}{m_Z^2} + \frac{\cos^2 \theta_{2}'}{m_{Z'}^2} \right) 
		(\Delta \mathcal{Q})^2
		\left( V_{tb}^\ast V_{ts} \right)^2 \Big/ N_{B\bar{B}}, \\
		\Delta C_\text{SLL}^1 = \Delta C_\text{SRR}^1 &= -\frac{1}{v^2} \left( \frac{\cos^2 (\alpha - \beta)}{m_h^2} + \frac{\sin^2 (\alpha - \beta)}{m_H^2} - \frac{[U_\gamma^{21}]^2}{m_{H_A}^2} \right) \left( 2 + \frac{1}{\tan^2 \beta} + \tan^2 \beta \right) \\
		& \hspace{0.5cm} \times m_s^2 \left( V_{tb}^\ast V_{ts} \right)^2 \Big/ N_{B\bar{B}}, \\
		\Delta C_\text{LR}^2 &=- \frac{2}{v^2} \left( \frac{\cos^2 (\alpha - \beta)}{m_h^2} + \frac{\sin^2 (\alpha - \beta)}{m_H^2} + \frac{[U_\gamma^{21}]^2}{m_{H_A}^2} \right) \left( 2 + \frac{1}{\tan^2 \beta} + \tan^2 \beta \right) \\
		& \hspace{0.5cm} \times m_s^2 
		\left( V_{tb}^\ast V_{ts} \right)^2 \Big/ N_{B\bar{B}},
	\end{aligned}
\end{equation}
with $\Delta \mathcal{Q}\equiv \left( \mathcal{Q}_{t_R} - \mathcal{Q}_{u_R} \right)$.
The  non-perturbative matrix elements can be parameterized by bag parameters, which are
detailed in Appendix \ref{app:pheo_detail}. In Table \ref{tab:BMU_basis}, a summary of
numerical values of bag parameters at  $\mu_b = m_b$ have been presented. 

For completeness, here we also exhibit the SM matrix element that 
\begin{equation}
	\langle B^0|H_{\rm eff}^{B\bar{B}}|\bar{B}^0\rangle_{\rm SM}=\frac{4}{3}N_{B\bar{B}}\left[\alpha_s^{(5)}(\mu_b)\right]^{-\frac{6}{23}}\left[1+\frac{\alpha_s^{(5)}(\mu_b)}{4\pi}J_5\right]m_{B_s}\eta_Bf_{B_s}^2B_{1}S_0(x_t),
\end{equation}
where QCD factor is $ \eta_B=0.5510 $, $ J_5=1.627 $ in the NDR scheme\cite{Buras:2001ra} and loop function $ S_0(x_t) $ is extracted form \cite{FermilabLattice:2016ipl}.
The evolution of Wilson coefficients, running from $ \Lambda $ to $ \mu_b $,
can be depicted by the following evolution matrix \cite{Buras:2001ra},
\begin{equation}\label{eq:U_m12}
	U(\mu_b,\Lambda)=
	\begin{pmatrix}
		[\eta(\mu_b,\Lambda)]_{\text{VLL}}	&0	&0	\\
		0	&[\eta(\mu_b,\Lambda)]_{\text{LR}}	&0	\\
		0	&0	&[\eta(\mu_b,\Lambda)]_{\text{SLL}}	\\
	\end{pmatrix}.
\end{equation}
The VRR and SRR sectors share the common as the above VLL and SLL ones.
 Numerical values of related parameters have been summarized in Table \ref{tab: BBmixing_RG}. 

\subsection{Electroweak and Higgs physics}
Various high-precision measurements around the $Z$ pole ($\sqrt{s} \approx m_Z$) provide an excellent testing ground for new physics (NP). Most of these measurements were conducted by LEP1 and SLC~\cite{Janot:2019oyi,ALEPH:2005ab,SLD:2000leq,SLD:2000ujp,SLD:2000jop,SLD:1996gjt} with remarkable precision, and they are summarized in Table~\ref{tab:Z_pole_obs}. Rather than elaborating on their extensive history, we refer interested readers to the cited references and instead focus on their connection to new physics through
\begin{equation}
    \begin{aligned}
        g_v^f &= I_3^f - Q_f\sin^2\theta_W, \quad
        g_a^f = I_3^f,\\
        \delta g_v^f &= \frac{g'}{2}\sin\theta_2'(\mathcal{Q}_{fR} + \mathcal{Q}_{fL})\big/N_g, \quad
        \delta g_a^f = \frac{g'}{2}\sin\theta_2'(\mathcal{Q}_{fR} - \mathcal{Q}_{fL})\big/N_g,
    \end{aligned}
\end{equation}
\begin{equation}
	\begin{aligned}
		G_f &\equiv \left(g_v^f\right)^2 + \left(g_a^f\right)^2, \quad
		\delta G_f \equiv 2\left(g_v^f \delta g_v^f + g_a^f \delta g_a^f\right), \\
		\delta\mathcal{F}_f &=\sum_fN_c^f\delta G_f, \quad
		\delta\mathcal{F}_q=\sum_qN_c^q\delta G_q, \quad
		\mathcal{F}_f=\sum_fN_c^f G_f, \quad
		\mathcal{F}_q=\sum_qN_c^q G_q,
	\end{aligned}
\end{equation}
where the prefactor is $N_g=e\cos\theta_2'/(2\sin\theta_W\cos\theta_W)$. The corresponding $Z$-pole formulas are given below~\cite{Ciuchini:2013pca}:
\begin{equation}\label{eq:gammaZ_np}
	\delta \Gamma_{Z,\text{NP}} = \frac{\alpha m_Z}{12c_W^2s_W^2}\delta\mathcal{F}_f, \quad
    \delta \Gamma_{Z,\text{NP}}^f = \frac{\alpha m_Z}{12c_W^2s_W^2}N_c^f\delta G_f,
\end{equation}
\begin{equation}\label{eq:sigma0_np}
	\delta \sigma_{\text{had,NP}}^0 = \frac{12\pi}{m_Z^2}\frac{G_e\mathcal{F}_q}{\mathcal{F}_f^2}\left(
	\frac{\delta G_e}{G_e}+\frac{\delta\mathcal{F}_q}{\mathcal{F}_q} - 2\frac{\delta\mathcal{F}_f}{\mathcal{F}_f}
	\right),
\end{equation}
\begin{equation}\label{eq:Af_np}
	\delta A_{f,\text{NP}} = -\frac{2\left[(g_v^f)^2 - (g_a^f)^2\right]}{G_f^2}\left(g_a^f\delta g_v^f - g_v^f\delta g_a^f\right),
\end{equation}
\begin{equation}\label{eq:AFB_np}
	\delta A_{\text{FB,NP}}^{0,f} = -\left[
	\frac{3g_v^fg_a^f\left[(g_v^e)^2 - (g_a^e)^2\right]}{G_fG_e^2}\left(g_a^e\delta g_v^e - g_v^e\delta g_a^e 
	+ e\leftrightarrow f\right)
	\right],
\end{equation}
\begin{equation}\label{eq:Rf_np}
	 \delta R_{\ell,\text{NP}}^0 = N_c^\ell\left(\frac{\delta \mathcal{F}_q}{G_\ell} - \frac{\mathcal{F}_q\delta G_\ell}{G_\ell^2}\right), \quad
	\delta R_{c,\text{NP}}^0 = N_c^q\left(\frac{\delta G_c}{\mathcal{F}_q} - \frac{G_c\delta \mathcal{F}_q}{\mathcal{F}_q^2}\right).
\end{equation}
With these formulas in hand, we can systematically construct the corresponding $\chi^2_{\rm EWPOs}$. 
Additionally, the formulas for SM predictions and the effective weak mixing angles $\sin\theta_{\rm eff}^{b,\ell}$ and $\sin\theta_{\rm eff}^{\rm Up, Down}$ used in our numerical analysis can be found in~\cite{Dubovyk:2019szj,Awramik:2006uz}. These results have been adopted and rigorously benchmarked by the ZFITTER group and its successors~\cite{Arbuzov:2005ma,Arbuzov:2023afc}.

Beyond the $Z$-pole observables, the $h f\bar{f}$ vertices are also constrained by current flavor-conserving Higgs experiments~\cite{CMS:2022dwd,ATLAS:2022vkf}. These constraints are typically derived by performing a fit to the signal strengths, defined as
\begin{equation}
    \mu_{if} \equiv \frac{\sigma_i\cdot\mathcal{B}_f}{\sigma_i^{\rm SM}\cdot\mathcal{B}_{f}^{\rm SM}},
\end{equation}
where $i$ represents a given initial production mode,\footnote{Here, we follow the CMS assumption $\sigma_i = \sigma_i^{\rm SM}$ when extracting $\mu_f$~\cite{CMS:2022dwd}.} and 
$f = b\bar{b},~c\bar{c},~\tau\bar{\tau},~\mu\bar{\mu}$ corresponds to various decay channels.
Under the Born approximation~\cite{Resnick:1973vg,Ellis:1975ap}, the partial width of Higgs boson decays into fermion pairs can be simplified as:
\begin{equation}
    \frac{\Gamma_{hff}^{\rm NP}}{\Gamma_{hff}^{\rm SM}} =
    \left| \frac{g_{hff}^{\rm NP}}{g_{hff}^{\rm SM}} \right|^2 =
    \left| \sin(\alpha - \beta) - \cos(\alpha - \beta) \frac{N_f}{m_f} \right|^2.
\end{equation}
We do not include constraints from light quark or lepton Yukawa couplings due to the lack of available measurements. However, we incorporate the correlation matrix from CMS~\cite{CMS:2022dwd} in the construction of $\chi^2_{\rm Higgs}$.

\section{Numerical Analysis}
\label{sec: num}	

In this section, we perform a numerical analysis of the FG2HDM. Our primary objective is to efficiently identify viable solutions after extending the particle spectrum and interactions beyond the SM. In the following analysis, we explicitly examine the parameter space for the two U(1)$'$ charges $\mathcal{Q}_{d_R}$ and $\mathcal{Q}_{\mu_R}$, along with $g'$, $\sin\theta_{2}'$, $s_{\alpha\beta}$
\footnote{Here, $s_{\alpha\beta}$ denotes $\sin(\alpha - \beta)$.}, $\tan\beta$, and the masses of four new particles: $m_{Z'}$, $m_H$, $m_{H^\pm}$, and $m_{H_A}$, with a focus on the aforementioned flavor observables.

\subsection{Input Parameters and Experimental status}\label{sub: num_input}

With more data  accumulated on the decay related to $b \to c \ell \nu$ since 
the first report of $R_{D^{(*)}}$ anomaly in 2012,
tensions still exist up to now. For the $R_{D^{(\ast)}}$ problem, there still retains a discrepancy of 3.31$\sigma$ after recent updates \cite{Belle-II:2024ami,HFLAV:moriond24}. Two additional lepton flavor universality (LFU) observables, $R_{J/\Psi} \equiv \mathcal{B}(B_c \to J/\Psi \tau \bar{\nu})/\mathcal{B}(B_c \to J/\Psi \mu \bar{\nu})$ and $R_{\Lambda_c} \equiv \mathcal{B}(\Lambda_b \to \Lambda_c \tau \bar{\nu})/\mathcal{B}(B_c \to \Lambda_c \mu \bar{\nu})$, supplement the $b \to c \tau \bar{\nu}$ current \cite{LHCb:2017vlu,LHCb:2022piu,Riti:2024lrs}. For the $\tau$ longitudinal-polarization asymmetry $P_\tau^{D^\ast}$ \cite{Tanaka:2012nw} in $D^\ast$ and the fraction of the $D^\ast$ longitudinal mode $F_L^{D^\ast}$ \cite{Tanaka:2012nw}, the latest observations are maintained by BELLE \cite{Belle:2016dyj} and LHCb \cite{LHCb:2023ssl}, respectively. 
All the observations related to $b \to c \ell \nu$, except  $R_{D^{(\ast)}}$,   are within  
$2 \sigma$ errors of the SM predictions. Besides, we do not include $P_\tau^D$ and $R_{\Upsilon(3S)}$ in the numerical analysis, due to the lack of experimental data for $P_\tau^D$ and the absence of scalar Wilson coefficients for $R_{\Upsilon(3S)}$ as reported in Ref. \cite{Iguro:2024hyk}. 
 The latest experimental status of aforementioned observables is summarized 
 in Table \ref{tab:FCCC_p}.

\begin{table}[b!]
	\centering
	\caption{\label{tab:FCCC_p}
	Summary of current experimental values and SM predictions related to $ b\to c\ell\nu $ process. The correlation between $ R_D$ and $R_{D^\ast} $ is -0.39 after Moriond 2024. 
	}
	\begin{tabular}{ccccccc}
		\hline\hline
		\makebox[0.06\textwidth][c]{}&
		\makebox[0.145\textwidth][c]{$R_D $}&
		\makebox[0.145\textwidth][c]{$R_{D^\ast}$}&
		\makebox[0.145\textwidth][c]{$P_{\tau}^{D^\ast}$}& 
		\makebox[0.145\textwidth][c]{$F_{L}^{D^\ast} $}&
		\makebox[0.145\textwidth][c]{$R_{J/\Psi}$}&
		\makebox[0.145\textwidth][c]{$R_{\Lambda_c} $}\\
		\hline
		SM 
		& 0.302(8)\cite{Duan:2024ayo} 
		& 0.257(5)\cite{Duan:2024ayo}
		& -0.518(6)\cite{Duan:2024ayo} 
		& 0.430(7)\cite{Duan:2024ayo} 
		& 0.258(4)\cite{Harrison:2020nrv,Duan:2024ayo} 
		& 0.332(10)\cite{Duan:2024ayo}\\
		Expt.
		& 0.342(26)\cite{HFLAV:2022esi} 
		& 0.287(12)\cite{HFLAV:2022esi} 
		& -0.38$ (^{+53}_{-55}) $\cite{Belle:2016dyj} 
		& 0.49(5)\cite{Iguro:2024hyk} 
		& 0.61(18)\cite{Iguro:2024hyk}
		& 0.271(72)\cite{Iguro:2024hyk,Bernlochner:2022hyz}\\
		\hline\hline
	\end{tabular}
\end{table}

\begin{table}[t]
	\centering
	\caption{\label{tab:BMU_basis}
	The averaged bag parameters $f_{B_s}^2B_{i}$
	(in the unit of $ \text{GeV}^2 $) adopted in the calculation of $ B_s^0- \bar{B}_s^0$ mixing, 
	at $ \mu_b=\bar{m}_b $ and in the $ \overline{\text{MS}}-\text{NDR} $ scheme \cite{DiLuzio:2019jyq}.}
	\begin{tabular}{cccccc}
		\hline\hline
		\makebox[0.08\textwidth][c]{i}&
		\makebox[0.13\textwidth][c]{1}&
		\makebox[0.13\textwidth][c]{2}& 
		\makebox[0.13\textwidth][c]{3}&
		\makebox[0.13\textwidth][c]{4}&
		\makebox[0.13\textwidth][c]{5}\\
		\hline
		$ f_{B_s}^2B_{i}$  & 0.0452(14) & 0.0441(17) & 0.0454(27) & 0.0544(19)&0.0507(17)\\
		\hline\hline
	\end{tabular}
\end{table}

\begin{table}[h]
	\centering
	\caption{\label{tab: BBmixing_RG}
	The inputs for evaluating evolution matrix $U(\mu_b,\Lambda)$ in
	Eq. \eqref{eq:U_m12},   with
	$\Lambda = 100\,\text{GeV}$ at $ \mu_b $ scale and 
	$\alpha_s$ is taken as $ \alpha_s^{(5)}=0.1179 $ \cite{Buras:2001ra}.
	}
	\begin{tabular}{ccc ccc ccc}
		\hline\hline
		\makebox[0.09\textwidth][c]{$[\eta]_{\text{VLL}}$}&
		\makebox[0.09\textwidth][c]{$[\eta_{11}]_{\text{LR}}$}&
		\makebox[0.09\textwidth][c]{$[\eta_{12}]_{\text{LR}}$}& 
		\makebox[0.09\textwidth][c]{$[\eta_{21}]_{\text{LR}}$}&
		\makebox[0.09\textwidth][c]{$[\eta_{22}]_{\text{LR}}$}&
		\makebox[0.09\textwidth][c]{$[\eta_{11}]_{\text{SLL}}$}&
		\makebox[0.09\textwidth][c]{$[\eta_{12}]_{\text{SLL}}$}&
		\makebox[0.09\textwidth][c]{$[\eta_{21}]_{\text{SLL}}$}&
		\makebox[0.09\textwidth][c]{$[\eta_{22}]_{\text{SLL}}$}
		\\
		\hline
		0.855 	& 0.926  &
		-0.037 	& -0.784 & 
		2.109 	& 1.588  & 
		1.820 	& -0.006 & 
		0.577 \\	
		\hline\hline
	\end{tabular}	
\end{table}

The $R_{K^{(\ast)}} $ released in December 2022 \cite{LHCb:2022vje} were finally found to be compatible with SM prediction, indicating the long expected 
LFU anomalies in $b\to s \ell^+ \ell^-$ have faded way.  However, NP opportunity has 
not exculded in the binned branching fractions and  particular component of angular distributions.
All the  binned global fit works \cite{Greljo:2022jac,Ciuchini:2022wbq,Wen:2023pfq,Altmannshofer:2023uci,Hurth:2023jwr,Bordone:2024hui}  in recent yeas  exhibit the common behaviors that allowing a deviation of $C_9$ can accommodate data.
 Recently, the unbinned amplitude analysis with improved Breit–Wigner function 
 \cite{LHCb:2023gel,LHCb:2023gpo,LHCb:2024onj} also show a deviation of
 $C_9$ (without a discrimination of lepton flavor species) from SM is about
  $ 2\sigma $. 
Incorporating around $200$ observables related to $b\to s\ell^+ \ell^-$, including
latest $R_{K^{(*)}}$ values,
the recent model independent global fit of NP parameters \cite{Wen:2023pfq}
indicates that
 \begin{equation}
	\begin{aligned}
		\Delta C_9^\mu&=-0.872\pm0.215,~&\Delta C_9^e&=-1.511^{+0.561}_{-0.533},\\
		\Delta C_{10}^\mu&=0.171^{+0.157}_{-0.175},~&\Delta C_{10}^e&=0.383^{+0.840}_{-0.424},
	\end{aligned}
\end{equation}
and 
\begin{equation}
	\begin{aligned}
		C_{S}^{\mu}& = 0.009^{+0.858}_{-0.845},~ 		
		C_{S}^{e} = -0.806^{+1.900}_{-1.238},
		&C_{S}^{\prime\mu}& = 0.012^{+0.858}_{-0.862},~ 	
		C_{S}^{\prime e}=-0.803^{+1.861}_{-1.194};
		\\
		C_{P}^{\mu}&=0.124^{+0.902}_{-0.910},~ 		
		C_{P}^{e}=-1.837^{+1.376}_{-0.930},
		&C_{P}^{\prime\mu}&=0.038^{+0.894}_{-0.913},~ 	
		C_{P}^{\prime e}=-1.652^{+1.200}_{-0.979}.
	\end{aligned}
\end{equation}
in its  S-IV scenario involving as general as  20 Wilson coefficients. The around $4 \sigma$
and $3 \sigma$
deviations in $C_9^\mu$ and $C_9^e$, respectively, as well as 
other Wilson coefficients  will be taken as inputs of
numerical analysis in this work.

For the consideration of NP in $B_s^0-\bar{B}_s^0$ system, we take 
$\Delta M_s$ as the typical observable in this analysis.
Combing the recent updated PDG value \cite{ParticleDataGroup:2022pth} 
with an inclusion of the latest value measured by LHCb  \cite{LHCb:2023sim}
\begin{equation}
		\Delta M_s^{\text{PDG}}=17.765\pm0.005~\text{ps}^{-1},
\end{equation}
as well as theoretical prediction in SM 
$\Delta M_s^{\text{SM}}=18.23\pm0.63~\text{ps}^{-1}$ \cite{Albrecht:2024oyn},
NP effect can be discriminated from the SM and parameterized \cite{DiLuzio:2019jyq,DiLuzio:2017fdq}  as
\begin{equation}
	\begin{aligned}
		\Big|1+\frac{M_{12}^{\text{NP}}}{M_{12}^{\text{SM}}}\Big|\sim\frac{\Delta M_s^{\text{exp}}}{\Delta M_s^{\text{SM}}}=0.974\pm0.034,
	\end{aligned}
\end{equation}
leading to
\begin{equation}
	\begin{aligned}
		-0.094 \leq \frac{M_{12}^{\text{NP}}}{M_{12}^{\text{SM}}} \leq 0.042
	\end{aligned}
	\label{eq:mixing-Ms}
\end{equation} 
within $2\sigma $ standard deviations. The range in Eq. \eqref{eq:mixing-Ms},
together with Table \ref{tab:BMU_basis}  and \ref{tab: BBmixing_RG},
serve as experimental and theoretical respective inputs in the following numerical calculations
of $B_s$ mixing system.

\begin{table}[t]
	\centering
        \caption{\label{tab:h_sigs}
        The measurements on signal strength  extracted from various decay channels 
        with the assumption $\sigma_i=\sigma_i^{\rm SM}$, in which Gaussian error approximation
        has been adopted \cite{CMS:2022dwd,CMS:2022fxs}. 
        }
        \begin{tabular}{ccccc}
		\hline\hline
		\makebox[0.08\textwidth][c]{$f$}&
		\makebox[0.13\textwidth][c]{$b\bar{b}$}&
		\makebox[0.13\textwidth][c]{$c\bar{c}$}& 
		\makebox[0.13\textwidth][c]{$\tau\bar{\tau}$}&
		\makebox[0.13\textwidth][c]{$\mu\bar{\mu}$}\\
		\hline
		$ \mu_f$  & 1.05(22) & 9.4(203)  & 0.85(10)& 1.21(45)\\
		\hline\hline
	\end{tabular}
\end{table}

Much more has been learned about the properties of the Higgs boson in the decade since its discovery.
In this work, we focus on its fermionic decay modes, which are of particular interest in
the following numerical analysis.
The latest results, based on partial Run-2 datasets from CMS~\cite{CMS:2022dwd, CMS:2022fxs} and ATLAS~\cite{ATLAS:2022vkf}, are summarized in Table~\ref{tab:h_sigs}.

Extensive measurements of EW observables have been conducted over decades, providing a robust foundation for precision tests of the SM and searches for NP. In this section, we present a systematic compilation of key observables near the 
$Z$-pole, including their experimental values, SM theoretical predictions, and corresponding NP corrections in the working model of current paper. These quantities  organized  in Table~\ref{tab:Z_pole_obs}, serve as inputs in the subsequent analysis.


\begin{table}[htbp]
	\caption{
Principal $Z$-pole observables, their SM predictions, and NP formulas. Measurements not separated by horizontal lines are correlated through correlation matrices\cite{ALEPH:2005ab}. The first set of $A_e$ and $A_\tau$ measurements are extracted from the combination of leptonic polarization and left-right asymmetry at SLD, while the second set follows from the LEP-1 experiments.
	}
	\label{tab:Z_pole_obs}
	\begin{center}
		\setlength{\tabcolsep}{15pt}
		\renewcommand\arraystretch{1.00}
		\begin{tabular}{lccc}
			\hline\hline
			Observables & Measurements & SM predictions & NP corrections  \\
			\hline
			$m_Z\,[\text{GeV}]$&$91.1876(21)$\cite{ParticleDataGroup:2022pth}&$91.1884$&\\
			$\Gamma_Z \, [\text{GeV}]$ & $2.4955(23)$ \cite{ALEPH:2005ab,Janot:2019oyi}  
			&$2.4946$ 
			&Eq. \eqref{eq:gammaZ_np}\\
			$\sigma_{\text{had}}^0 \, [\text{nb}]$ & $41.4802(325)$ \cite{ALEPH:2005ab,Janot:2019oyi}  &$41.4922$ &Eq. \eqref{eq:sigma0_np} \\
			$R_e$ & $20.804(50)$ \cite{ALEPH:2005ab} &$20.733$ 
			& \\
			$R_{\mu}$ & $20.785 (33)$ \cite{ALEPH:2005ab} &$20.733$ 
			&Eq. \eqref{eq:Rf_np} \\
			$R_{\tau}$ & $20.764 (45)$ \cite{ALEPH:2005ab} &$20.780$ 
			& \\
			$A_{\text{FB}}^{0,e}$ & $0.0145 (25)$ \cite{ALEPH:2005ab} &$0.0163$ 
			& \\
			$A_{\text{FB}}^{0,\mu}$ & $0.0169 (13)$ \cite{ALEPH:2005ab} &$0.0163$ 
			&Eq. \eqref{eq:AFB_np} \\
			$A_{\text{FB}}^{0,\tau}$ & $0.0188 (17)$ \cite{ALEPH:2005ab} &$0.0163$ 
			& \\
			\hline
			$R_b$ & $0.21629 (66)$ \cite{ALEPH:2005ab} &$0.21584$ 
			& \\
			$R_c$ & $0.1721 (30)$ \cite{ALEPH:2005ab} &$0.17222$ 
			&Eq. \eqref{eq:Rf_np} \\
			$A_{\text{FB}}^b$ & $0.0996 (16)$ \cite{ALEPH:2005ab} &$0.1035$ 
			& \\
			$A_{\text{FB}}^c$ & $0.0707 (35)$ \cite{ALEPH:2005ab} &$0.0739$ 
			&Eq. \eqref{eq:AFB_np} \\
			$A_b$ & $0.923 (20)$ \cite{ALEPH:2005ab} &$0.935$ 
			& \\
			$A_c$ & $0.670 (27)$ \cite{ALEPH:2005ab} &$0.668$ 
			&Eq. \eqref{eq:Af_np} \\
			\hline
			$A_e$ & $0.1516 (21)$ \cite{ALEPH:2005ab} &$0.1476$ & \\
			$A_{\mu}$ & $0.142 (15)$ \cite{ALEPH:2005ab} &$0.1476$ 
			&Eq. \eqref{eq:Af_np} \\
			$A_{\tau}$ & $0.136 (15)$ \cite{ALEPH:2005ab} &$0.1476$ & \\
			\hline
			$A_e$ & $0.1498 (49)$ \cite{ALEPH:2005ab} &$0.1476$ & \\
			\hline
			$A_{\tau}$ & $0.1439 (43)$ \cite{ALEPH:2005ab} &$0.1476$ 
			&Eq. \eqref{eq:Af_np} \\
			\hline
			$A_s$ & $0.895 (91)$ \cite{ALEPH:2005ab} &$0.936$ & \\
			\hline
			$A_{\text{FB}}^{0,s}$ & $0.0976 (114)$ \cite{ALEPH:2005ab} &$0.1036$ 
			&Eq. \eqref{eq:AFB_np} \\
			\hline
			$R_{uc}$ & $0.166 (9)$ \cite{ParticleDataGroup:2022pth} &$0.1722$ 
			&Eq. \eqref{eq:Rf_np} \\
			\hline
		\end{tabular}
	\end{center}
\end{table}

In addition to the essential physical observables mentioned above and their associated parameters , Table \ref{tab:Input_para}  provides a summary of other basic input parameters required for the subsequent theoretical calculations.

\begin{table}[t]
	\caption{\label{tab:Input_para}
	Basic input parameters in the numerical analysis. 
	}
\begin{tabular}{cccc}
	\hline\hline
	\makebox[0.15\textwidth][c]{Parameters}&
	\makebox[0.15\textwidth][c]{Values}&
	\makebox[0.15\textwidth][c]{Parameters}&
	\makebox[0.15\textwidth][c]{Values}\\
	\hline
	$ m_{b} $& 4.18( $\!^{+3}_{-2} $)~GeV \cite{ParticleDataGroup:2022pth}& $ m_{t} $ & 173.50(30)~GeV \cite{ParticleDataGroup:2022pth} \\
	$ m_{c} $&1.27$ (2) $~GeV \cite{ParticleDataGroup:2022pth}&$ m_{s} $ &93$\left(^{+11}_{-5}\right)$~MeV \cite{ParticleDataGroup:2022pth}\\
	$ m_{d} $&4.67$ (^{+48}_{-17}) $~MeV \cite{ParticleDataGroup:2022pth}&$ m_{u} $&2.16$( ^{+49}_{-26} )$~MeV \cite{ParticleDataGroup:2022pth}\\
	$ m_{e} $&0.5109989461$ (31) $~MeV \cite{ParticleDataGroup:2022pth}&$ m_{\mu} $&105.6583745$ (24) $~MeV \cite{ParticleDataGroup:2022pth}\\
	$ m_{\tau} $&1776.86(12)~MeV \cite{ParticleDataGroup:2022pth}&$m_W$&80.377(12)~GeV \cite{ParticleDataGroup:2022pth}\\
	$ m_{Z} $&91.1876(21)~GeV \cite{ParticleDataGroup:2022pth}&$m_h$&125.12(17)~GeV \cite{ParticleDataGroup:2022pth}\\
	$ \bar{m_t}(\bar{m_t}) $&163.53(83)~GeV \cite{ParticleDataGroup:2022pth}&$ m_{\nu_i}$&$ <10^{-11} $ GeV\\
	$ m_{B_u} $&5279.34(12)~MeV \cite{ParticleDataGroup:2022pth}&
        $ \tau_{B_u} $&1.638(4) ~ps \cite{ParticleDataGroup:2022pth}\\
	$ f_{B_u} $&190.0(13)~MeV \cite{ParticleDataGroup:2022pth}&
        $ m_{B_s} $&5366.92(10)~MeV \cite{ParticleDataGroup:2022pth}\\
        $ G_F $&1.1663787(6)~GeV$ ^{-2} $ \cite{ParticleDataGroup:2022pth}&
        \\
	\hline
	$\alpha_s(m_Z)$&0.1179(9) \cite{ParticleDataGroup:2022pth}&$\alpha_{\text{e}}(m_{Z})$&1/127.944(14) \cite{ParticleDataGroup:2022pth}\\
	$ \sin^2\theta_W $&0.23121(4) \cite{ParticleDataGroup:2022pth}&$ S_0(x_t) $&2.322(18) \cite{FermilabLattice:2016ipl}\\
	\hline
	$ \lambda $&0.22500(67) \cite{ParticleDataGroup:2022pth}&$ A $&0.826$ (^{+18}_{-15}) $ \cite{ParticleDataGroup:2022pth}\\
	$\bar{\rho}$&$ 0.159(^{+10}_{-10}) $ \cite{ParticleDataGroup:2022pth}&$\bar{\eta}$&$ 0.348(^{+10}_{-10}) $ \cite{ParticleDataGroup:2022pth}\\
	\hline\hline
\end{tabular}
\end{table}


\subsection{Numerical Results}\label{sub: num_res}

The decays $b\to c \ell \nu$ and  
$b\to s \ell^+ \ell^-$  are respectively attributed
by charged current and neutral current interactions. 
Hence separated analyses are performed as the first step.
Although it appears that the two processes are of less correlated,
they can be connected in certain specific UV-completed NP models. 
Such a correlation within the framework of  the FG2HDM will be explored  below.

\begin{figure}[ht]
	\centering
	\includegraphics[width=0.9\linewidth]{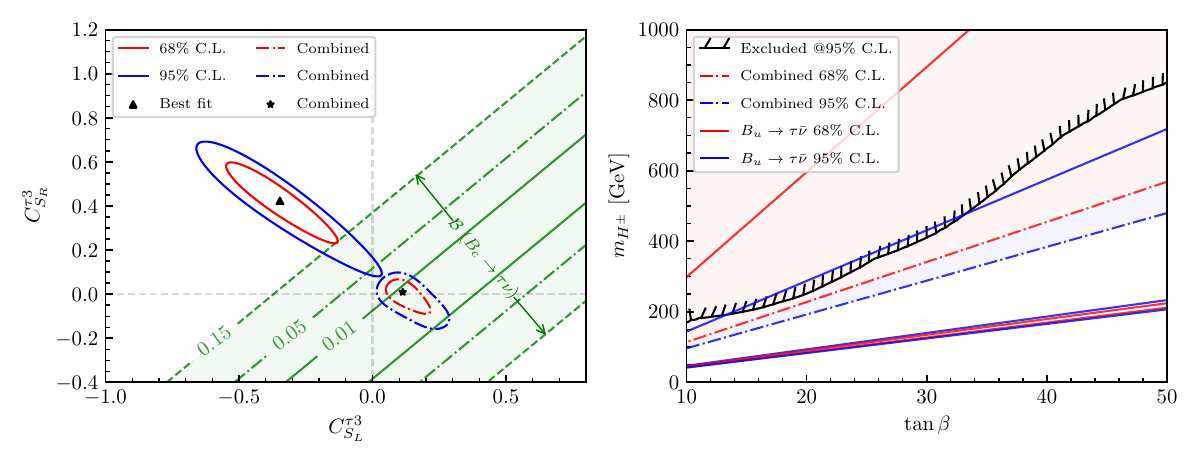}
	\caption{The allowed parameters space by $b\to c \ell \nu$ process: (Left) 
	model-independent  Wilson Coefficients $ C_{S_L}^{\tau3}$ and $  C_{S_R}^{\tau3} $; (Right) model-dependent allowed regions for $ \tan\beta $ as well as $ m_{H^\pm} $ where the excluded areas are extracted from ATLAS\cite{ATLAS:2018gfm} for reference.}
	\label{fig:FCCC_1}
\end{figure}

We first take $b\to c \ell \nu$ alone to put constraints on NP parameters in terms of
both model-independent and dependent ways as presented in Fig. \ref{fig:FCCC_1}.
In the left panel,  best fit of model-independent
coefficients $ C_{S_L}^{\tau3}$ and $ C_{S_R}^{\tau3}$ 
in the confidence level of $68\%$ and $95\%$ are exhibited 
on the top left, incorporating data in Table \ref{tab:FCCC_p}. 
The best fit point $( C_{S_L}^{\tau3},  C_{S_R}^{\tau3})=
{(-0.35,0.42)}$ indicates a negative $C_{S_L}^{\tau3}$ and positive $ C_{S_R}^{\tau3}$
are preferred by data at 95\% C.L.
This shows clearly why 2HDM-II is not preferred since the signs of the two  corresponding  
coefficients in 2HDM-II are identical.
Though not measured yet, a future measurement of decay $B_c\to \tau \nu$
will further discriminate parameter space. 
By conjecturing that $ \mathcal{B}(B_c\to\tau\nu)<15\% (5\%, 1\%)$, the solution region (dotdashed contours)
move to down right entirely, with a possibility to flip the sign of $C_{S_L}^{\tau3}$ at $2 \sigma$
level. 
The magnitude 
of the coefficients also decreases,  leading to even larger charged Higgs mass
and relative smaller $\tan\beta$ as shown with the dashed contours in 
the left panel as well. 
This change can be well interpreted from the allowed region of $ \mathcal{B}(B_c\to\tau\nu)$
marked in light green in the same plot. 
It is noticed that the measured branching ratio $\mathcal{B}(B_u\to\tau\nu)_{\rm exp}=(1.09\pm0.24)\times10^{-4}$\cite{ParticleDataGroup:2022pth} is indeed smaller
than that of the $B_c\to\tau\nu$ process, as assumed above. 
The right panel of Fig.~\ref{fig:FCCC_1} illustrates that the decay process $B_u \to \tau \nu$ can offer complementary information within the framework of the FG2HDM. 
Especially, 
when compared to the combined constraints from the $b \to c \ell \nu$ transitions,
it
provides even stronger bounds on the parameter space of $(\tan\beta, m_{H^\pm})$.
We also put a direct search bound provided by ATLAS \cite{ATLAS:2018gfm} in the right panel, 
the corresponding recasting will be combined in a separate work.

\begin{figure}[t]
	\centering
	\includegraphics[width=0.9\linewidth]{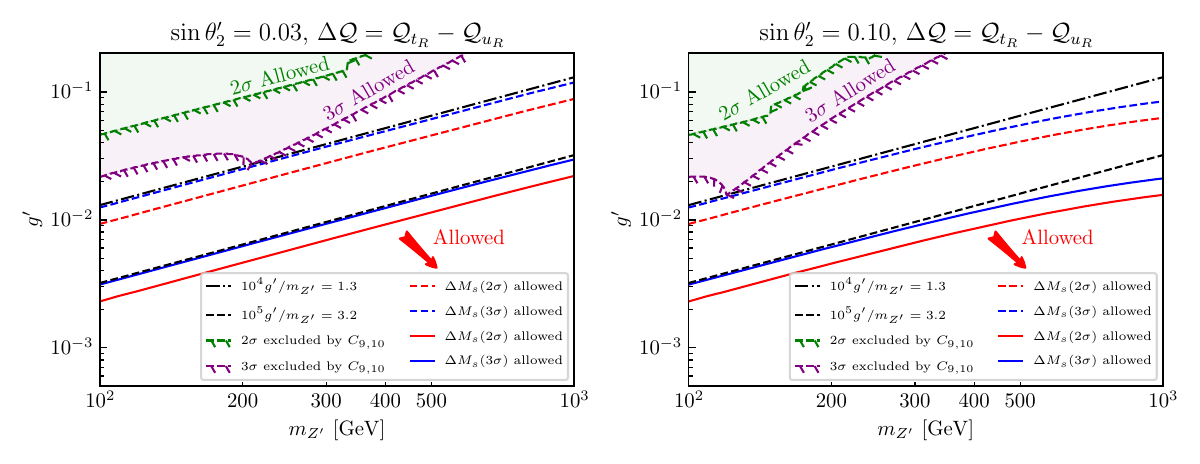}
	\caption{Solution space constrained by the new $ Z' $ boson mediated $ b\to s\ell\ell $ and $ B_s$-$ \bar{B}_s $ mixing: region with green(purple) shade represents the 2(3)$ \sigma $ survival area that provided by $ C_{9,10} $. Boundary with Red(blue) and its filling side represents the 2(3)$ \sigma $ requirement of $ \Delta M_{s} $ at fixed $ \Delta\mathcal{Q}$. Solid(dashed) boundary represents $ \Delta\mathcal{Q}=4(1)  $. The value of $ \sin\theta_{2}' $ holds 0.03(0.10) in the left(right) panel.
	}
	\label{fig:m12zp}
\end{figure}

The  $b\to s$ observables receive contributions from
both vector and scalar operators generated from
neutral gauge bosons and neutral scalars.
Here we firstly  focus on $Z^\prime$ sector by combing the restrictions
on $\Delta C_{9,10}$ and $\Delta C_{VLL}$. 
As shown in
Eqs. \eqref{eq:C9C10} and \eqref{eq:del_F=2}, 
the interaction strength is determined by mixing angle $\sin\theta_{2}'$ and
two $U(1)^\prime$ charges $\mathcal{Q}_{d_R},\mathcal{Q}_{\mu_R}$.
Two conjectures on mixing, SM-like (with $\sin\theta_{2}'=0.03 $) 
 and SM-deviated (with $ \sin\theta_{2}'=0.10$), 
 are made to carry out  further numerical analysis.
 For the consideration of naturalness, the two free $U(1)^\prime$ 
 charges 
 are set to be within the range
$ \mathcal{Q}_{d_R},\mathcal{Q}_{\mu_R}\in[-1,1] $, leading to
$\Delta \mathcal{Q} \in [-4,4]$. 
In  practical calculation, we choose $\Delta \mathcal{Q}=4$ and $1$ 
as two typic test options.  The remaining task is then to capture the allowed
regions in $(g',m_{Z'})$ plane. The survival space allowed
by $b\to s \ell^+ \ell^-$ is obtained from the $\chi^2(C_9, C_{10})$ constructed
by a combination of $C_{9,10}^{e,\mu}$, while allowed region for $B_s^0$ mixing 
originates from Eq.  \eqref{eq:mixing-Ms}.
Numerical results for the two conjectures with specific charge assignments are presented in Fig. \ref{fig:m12zp}. Taking the SM-like conjecture (left panel of Fig. \ref{fig:m12zp}) as an example, the permitted areas for $b\to s\ell^+ \ell^-$ and $\Delta M_s$ occupy the upper left and lower right portions, respectively. The overlapping space of these areas represents the desired survival region. 
To provide complementary information by a comparison
with charge-mass ratio 
$g_1/m_Z=e/m_W=3.8\times 10^{-3} \,\text{GeV}^{-1}$ in SM, 
we also plot two lines
$g'/m_{Z'}=1.3 (0.32)\times 10^{-4}\,\text{GeV}^{-1}$
 in black  in each
panel of Fig. \ref{fig:m12zp}. 
Apparently, data favors small $\Delta \mathcal{Q}$ case since no
overlapping region exists for the   $\Delta \mathcal{Q}=4$ scenario
in the left panel.  Therefore, with the condition of
$ \Delta \mathcal{Q}\leq 1 $ and $ g'/m_{Z'}>1.3\times10^{-4} $
the  solution space for $(g',m_{Z'})$ will be allowedable.
In the conjecture of SM-deviated case, as shown in the right panel, 
the solution space could exist but with smaller $Z'$ mass, which
will receive more strict examination from collider physics. 

\begin{figure}[t]
	\centering
	\includegraphics[width=0.9\linewidth]{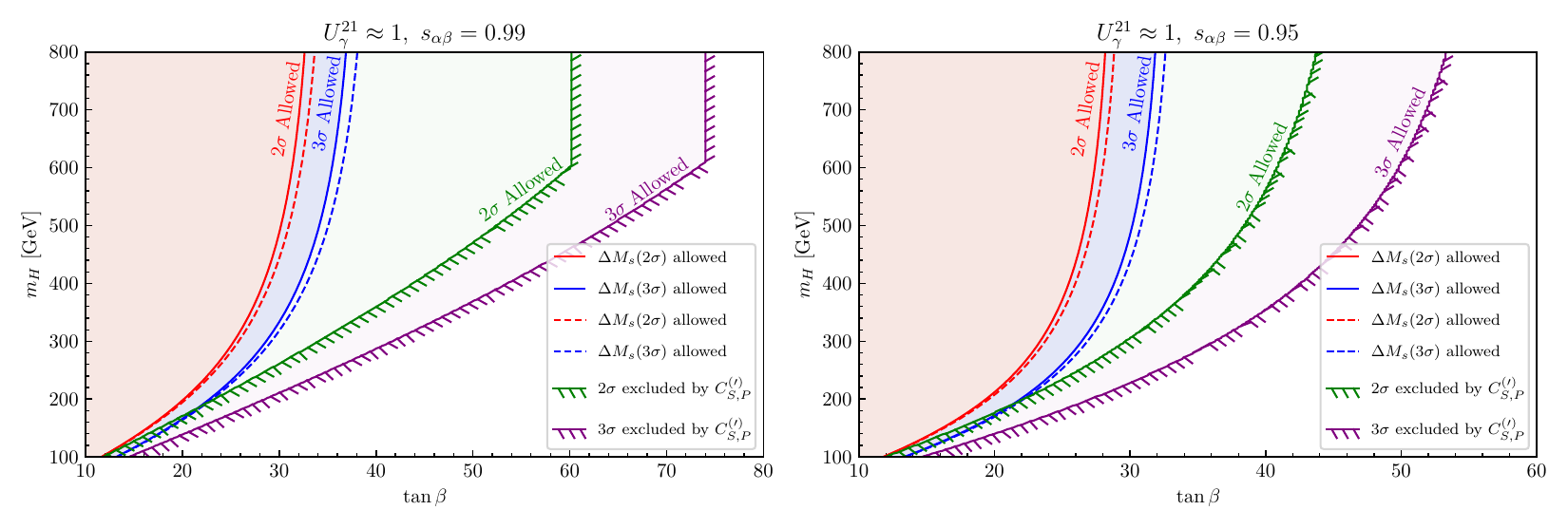}
	\caption{Solution space constrained by scalar boson mediated $ b\to s\ell\ell $ and $ B $-$ \bar{B} $ mixing: region with green(purple) shade represents the 2(3)$ \sigma $ allowed area provided by $ C_{S,P}^{(\prime)} $ at fixed value $ m_{H_A}=500$ GeV. Solid red(blue) boundary and its filling shade denotes the constraint from $ \Delta M_{s} $ mixing at fixed value $ m_{H_A}=500$ GeV. The dashed ones show the changes when $ m_{H_A} $ is fixed at 520 GeV.
	}
	\label{fig:m12h}
\end{figure}

\begin{figure}[b]
    \centering
    \includegraphics[width=0.9\linewidth]{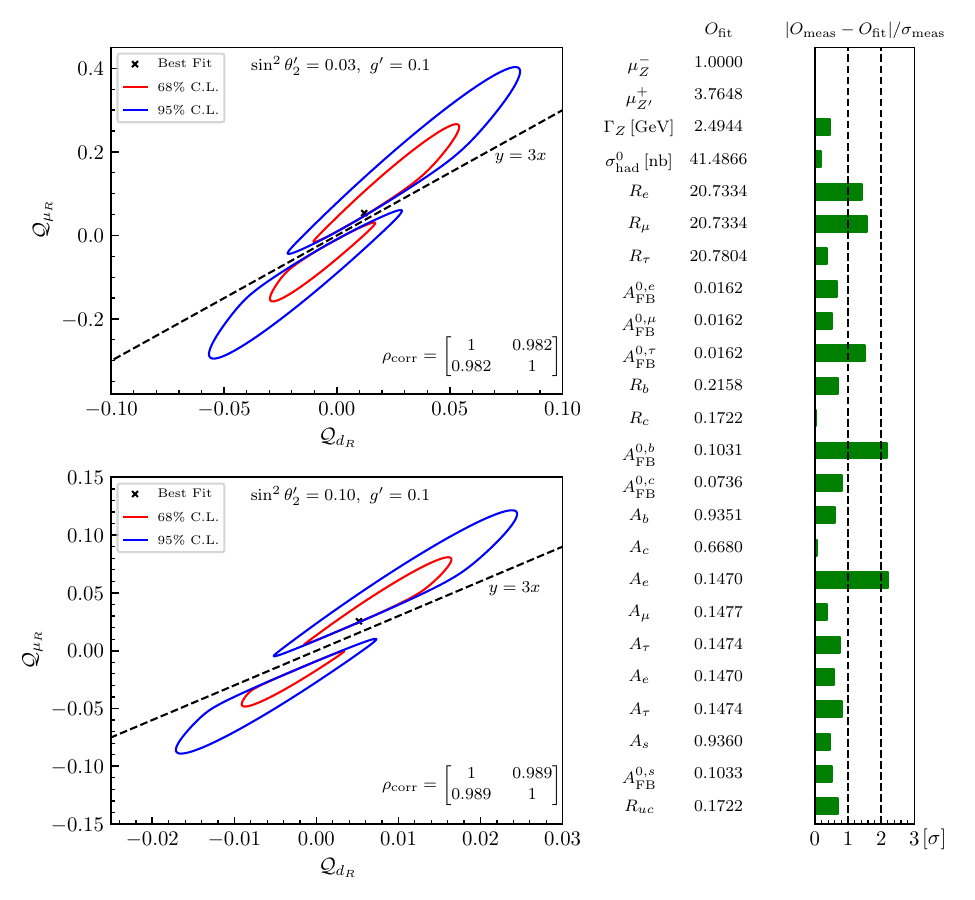}
    \caption{
\textbf{Left panel}: Solution space constrained by EWPOs at $Z$-pole. 
The red (blue) contour represents the boundary at the 68\% (95\%) confidence level.  
\textbf{Right panel}: Fit results and their deviations from experimental measurements. The values are obtained under the parameter settings:  $v_S = 4 \times 10^5$~GeV, and $\tan\beta = 15$
and two typical combinations of $g'$ and $\sin\theta_2'$.
    }
    \label{fig:Zpole_fit}
\end{figure}

The parameters  
$(m_{H}, m_{H_A}, \tan\beta, s_{\alpha\beta}, U_\gamma^{21})$
are related to scalar contribution to $b\to s$ processes 
from Eqs. \eqref{eq:CSCp} and \eqref{eq:del_F=2}.
Assuming a heavy scalar singlet as already adopted in this work,
the relevant $U_\gamma$ entry is close to unity, $U_\gamma^{21}\approx 1$.
The yet measured rotation between Higgs basis is conjectured as
$s_{\alpha\beta}=0.99$ and $0.95$, respectively. 
Hence a contribution from the observed neutral scalar $h$ is negligible. 
Without loss of generality, we fix the pseudoscalar mass as $m_{H_A} = 500\,\text{GeV}$
to explore the behavior of the remaining heavy neutral Higgs.
After setting the above parameters, we present numerical relations  between $\tan\beta$ and 
$m_H$   in Fig. \ref{fig:m12h}.
As illustrated in the left panel of  Fig. \ref{fig:m12h}, 
although large $\tan\beta$ is supported by 
 $\Delta C_{S,P}^{(\prime)}$ 
 constrained from $b\to s\ell^+\ell^-$,
$\Delta M_s$ with 
less constrained neutral scalar mass
prefers smaller $\tan\beta$. Hence the overlapping area, mainly originated from  
$\Delta M_s$,
suggests a strong preference for  $ \tan\beta<28 $, while the range of heavy quark mass 
$m_{H} $ remains broad, extending from hundreds of GeV to TeV as long as the $\tan\beta$ is sufficiently small. 
For the SM-deviated ($s_{\alpha\beta} = 0.95$) conjecture in
the right panel of Fig. \ref{fig:m12h}, the constraint from  
 $ C_{S,P}^{(\prime)} $ becomes more strict, while 
$ B_s $-$ \bar{B}_s $ mixing further suppresses 
the selections of value of $ \tan\beta $ similar to the situation in SM-like one.
Nevertheless, the solution space for SM-deviated conjecture can accommodate all the 
related data as well.

A comparison of the phenomenological results from flavor physics in Fig. \ref{fig:m12zp} with the phenomenological results from electroweak physics in Fig. \ref{fig:Zpole_fit} is both insightful and interesting. As we discuss in the narrative of Fig. \ref{fig:m12zp}, $B_s^0$ mixing provides us with the basic requirements $\Delta \mathcal{Q}<1$ and $g'/m_{Z'}>1.3\times10^{-4}$. When combined with the $g'=0.1$ setting in Fig. \ref{fig:Zpole_fit}, the value of $m_{Z'}$ is restricted to be less than 800 GeV. Furthermore, when combined with the limitations of $C_9$ and $C_{10}$, the range below 500 GeV is deemed to be more secure. 
Fig. \ref{fig:Zpole_fit} demonstrates refined $U(1)'$ charge assignments, and the key findings are as follows:
(1) The results of the $Z$-pole measurements suggest that viable solutions for constrained $U(1)'$ charges exist in both SM-like and SM-deviated scenarios, albeit within a narrow parameter space. Within these ellipse areas, the corresponding values of $\Delta \mathcal{Q}$ vary from -0.2 to 0.2, which are accommodated with the results derived from the flavor constraints in Fig. \ref{fig:m12zp}.
(2) The elements of correlation matrix and the auxiliary relation ($\mathcal{Q}{\mu_R} = 3\mathcal{Q}{d_R}$) indicate that the BGL symmetry remains compatible with current electroweak precision data.
(3) The right panel of Fig. \ref{fig:Zpole_fit} demonstrates that the best-fit point in a SM-like situation is consistent with the existing constraints from the $Z$-pole measurements. The trial solution $\mu_{Z^\prime}^+=3.76$ indicates that solutions which do not exceed the limits of flavor physics $m_{Z^\prime} < 500$ GeV can also exist.

\begin{figure}[t]
    \centering
    \includegraphics[width=0.9\linewidth]{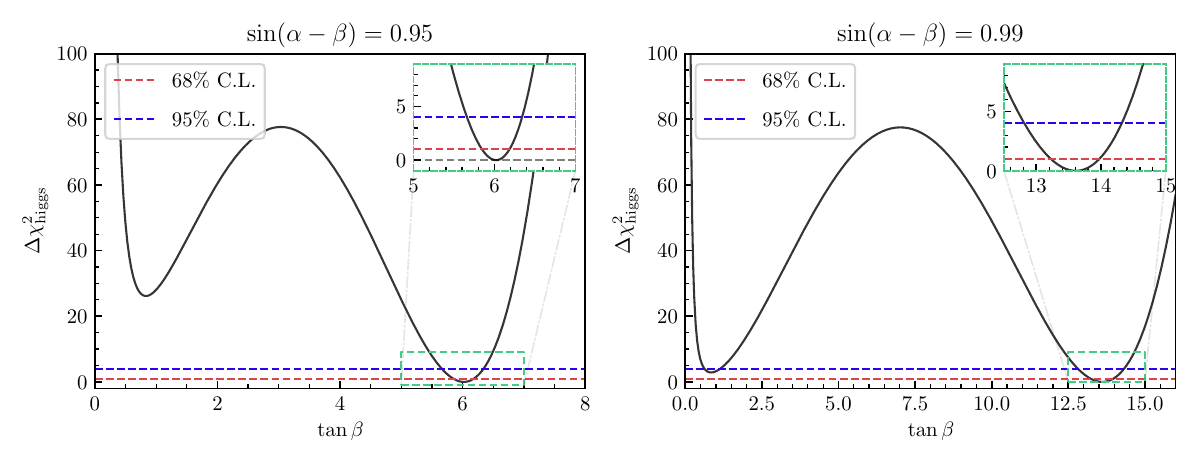}
    \caption{
	Profiles of $\Delta\chi^2 = \chi^2_{\rm Higgs} - \min(\chi^2_{\rm Higgs})$ in different scenarios.  
	The left (right) panel corresponds to the SM-deviated (SM-like) conjecture.  
	The dashed red (blue) line denotes the boundary at the 68\% (95\%) confidence level.  
	The insets in each subplot provide zoomed-in views around the respective minima.
    }
    \label{fig:hsig_fit}
\end{figure}

The constraints on the observed Higgs properties, derived from Fig. \ref{fig:hsig_fit}, reveal two key features. First, the value of 
$\tan\beta$  in the vicinity of the relatively large global minimum decreases as 
$s_{\alpha\beta}$  deviates further from unity.
Second, extrapolating the SM limit
$s_{\alpha\beta}\to 1$ suggests the existence of a local minimum with 
$\tan\beta<1$, which could arise in regions of the parameter space where Higgs couplings deviate significantly from Standard Model predictions. These trends highlight the interplay between 
$\tan\beta$ and the Higgs mixing angle 
$\alpha-\beta$  in shaping the phenomenological implications of the observed scalar sector.


\begin{figure}[t]
	\centering
		\includegraphics[width=1.0\linewidth]{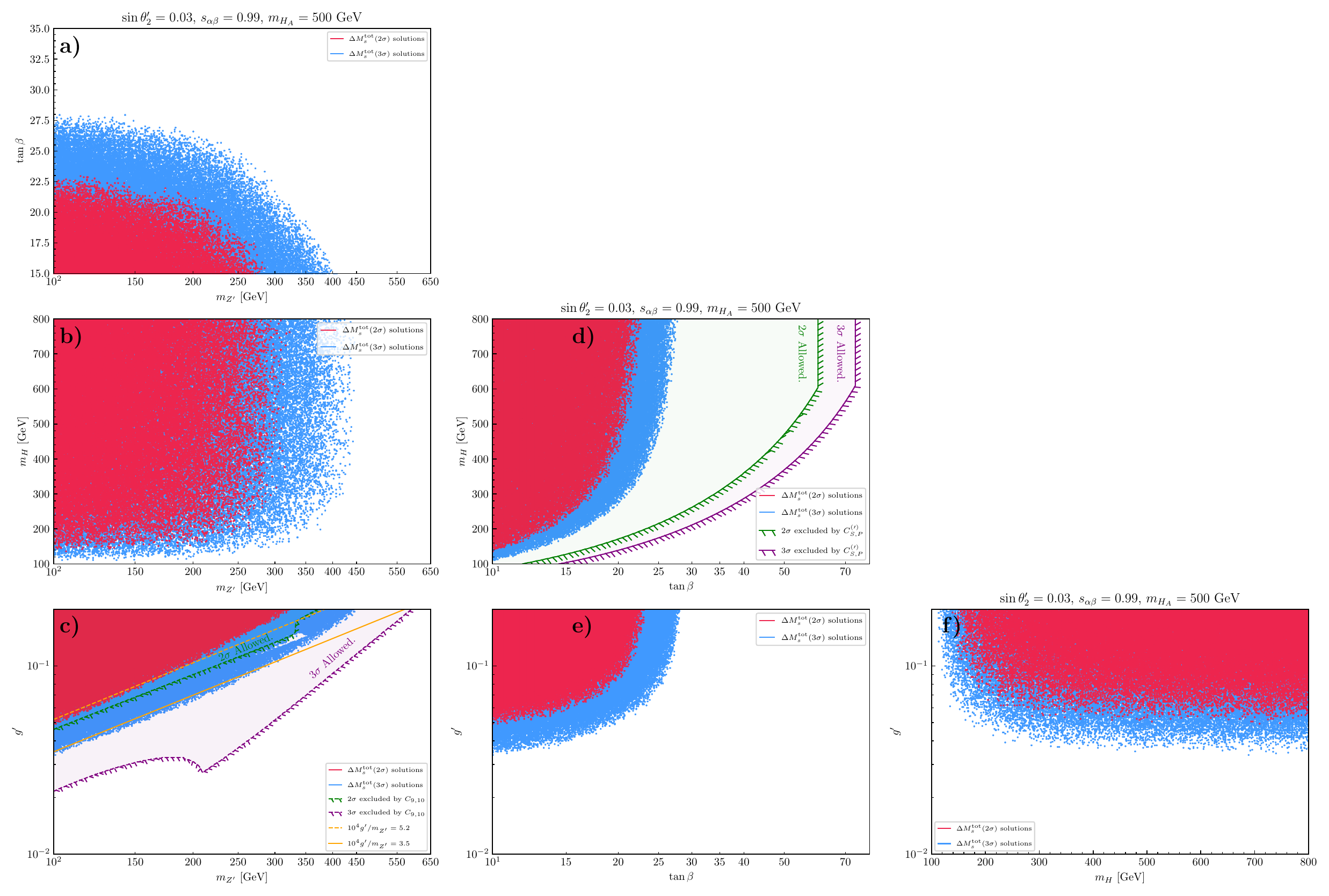}
	\caption{Solution space constrained by both scalar sector and gauge sector together.
	The regions with green(purple) shade have been explained in the fig.(\ref{fig:m12zp}) as well as in the fig.(\ref{fig:m12h}). Region formed by red(blue) data points indicates the projection of solutions space into the corresponding plane at the confidence level of 2(3)$ \sigma $.
	These solutions are obtained under the assumption of fixed $ \sin\theta_{2}'=0.03 $, $ s_{\alpha\beta}=0.99 $, and $ m_{H_A}=500 $ GeV.
	}
	\label{fig:m12tot}
\end{figure}

In the previous individual analyses (presented in Fig.\ref{fig:FCCC_1}, Fig.\ref{fig:m12zp} and Fig.\ref{fig:m12h}), correlations among parameters and observables have not yet been taken into account. Given that the FG2HDM parameter $\tan\beta$ influences both charged and neutral Higgs interactions, and that $b \to s$ observables receive contributions from both neutral Higgs and neutral gauge bosons, correlation effects may be significant. Therefore, we perform a combined analysis that incorporates measurements of $b \to c\ell \nu$, $b \to s \ell^+ \ell^-$, and $B_s^0$-$\bar{B}_s^0$ mixing, with all necessary FG2HDM parameters involved.

In  the SM-like conjecture with 
$\sin\theta'_2=0.03, s_{\alpha\beta}=0.99,  m_{H_A}=500\,\text{GeV}$, 
the higher dimensional global fit is carried out based on the $\chi^2$ function
 incorporating all previous measurements. With the best-fit point  
located at $\chi^2_{\text{min}}=26.69$, the $2\sigma$ and $3\sigma$ allowed region 
for $(m_{Z'}, m_H, \tan\beta, g')$
are plotted in red and blue as presented in  Fig. \ref{fig:m12tot}. 
We indeed  find a proper solution space within FG2HDM to accommodate flavor 
data studied in this work. Compared with constraints given by single observables,
parameter space have shifted due to the correlation effect. 
As displayed in
the left panel of Fig. \ref{fig:m12zp}, the mass of $Z'$ could be around 
$220\,\text{GeV}$, while in the combined analysis this value is extended to 
$450\,\text{GeV}$ shown in Fig.\ref{fig:m12tot}(a).
On the other hand,  the range for some parameters, like $\tan\beta$, has shrunk significantly
in the combined analysis displayed in Fig.\ref{fig:m12tot}(e).
This further implies that contributions to $b\to s$, from the gauge and scalar sectors,
are in direct competition with one another.  
Now we may safely conclude that typical model parameters 
$g', m_{Z'},  \tan\beta$ have been found to have their explicit boundaries
\begin{equation}
m_{Z'}<450\,\text{GeV}, 
\quad
\tan\beta< 28,
\quad  g'>10^{-2} 
\end{equation}
at $3\sigma$ level, while the masses of scalars are still 
less constrained.  

The ranges here obtained indirectly from flavor physics are expected to be
further improved by direct search of new particles in colliders, which will
be studied separately.


\section{Concluding Remarks}
\label{sec:con}

In this work, we systematically study an unique UV-completed two-Higgs doublet model (2HDM) extended by a scalar singlet. Imposed with a specific $U(1)$ flavor gauge symmetry, the Yukawa matrices are designed with special structures so that tree-level flavor-changing neutral current (FCNC) interactions, controlled by the CKM matrix, occur uniquely in the down-type quark sector. In the decoupling limit, the charged and neutral scalars, except the superheavy one from the singlet scalar, contribute to processes of interest. The additional neutral gauge boson $Z'$ arising from the gauged $U(1)$ flavor symmetry also induces tree-level FCNC processes in the down-type quark sector. 

We investigate the processes $b \to c \ell \nu$ and $b \to s \ell^+ \ell^-$, where some anomalies still exist, in conjunction with the related and well-measured observable in $B_s^0$-$\bar{B}_s^0$ mixing, EWPOs, and Higgs fermionic decays, to further explore the FG2HDM parameter space. We find that:

\begin{itemize}
    \item The FG2HDM has distinct features compared to the general 2HDM. For example, the identical signs of $C_{S_{L}}^{\tau 3}$ and $C_{S_{R}}^{\tau 3}$ generated in 2HDM-II cause tension between 2HDM-II and $R_{D^{(*)}}$. This tension can be alleviated by flipping one of the signs, which is realized in the FG2HDM by integrating the charged Higgs, as shown in Fig. \ref{fig:FCCC_1}.
    A combination of the measured branching fraction, $\mathcal{B}(B_u\to\tau\nu)$, and the conjecture regarding the $B_c\to \tau\nu$ process indicates that the charged Higgs mass must be heavy.
    \item The tree-level down-type quark FCNC interactions can be induced by both the heavy neutral scalar and the neutral gauge boson, with a  competition between each other. Constraints on individual sectors can be obtained from data, as presented in Fig. \ref{fig:m12zp} and Fig. \ref{fig:m12h}.
    \item 
    As illustrated in Fig. \ref{fig:Zpole_fit} and Fig. \ref{fig:hsig_fit}, 
    the current constraints imposed by the EW $Z$-pole observables and the Higgs physics
    can be simultaneously accommodated by setting a new tiny $U(1)'$ charge and establishing the value of $\tan\beta$ to be approximately $\mathcal{O}(10)$.
    \item Correlations established via common model parameters and physical observables shift the survival parameter space of a single observable. In the comprehensive combined analysis, we obtain allowed parameters in the SM-like conjecture.
 The upper bounds of the $Z'$  mass is found to be $450\,\text{GeV}$, while the upper bound for 
 $\tan\beta$ is $28$. On the other hand, the lower bound of the coupling constant of the new $U(1)$ interaction is $10^{-2}$, leading to a lower bound on the charge-mass ratio \( g'/m_{Z'} \) of around $3.5\times 10^{-4}$.
\end{itemize}

By simultaneously accommodating the flavor anomalies existing in both $b \to c \ell \nu$ and $b \to s\ell^+ \ell^-$ processes, the survival parameter space of the FG2HDM is found to be large. Preliminary explorations of the model, focusing on electroweak physics, as well as Higgs boson fermionic decay modes, have been conducted. Meanwhile, the obtained mass bound of $Z'$ requires a further examination from collider physics, which will be discussed in other place separately.

\acknowledgments
The authors benefited from helpful discussions with  Feng-Zhi Chen and Xin-Qiang Li. 
Discussions  with Junmou Chen and Mengchao Zhang  at the preliminary stage of the present work
 is also acknowledged. 
This work is supported by NSFC  under Grant  Nos.  12475095 and U1932104.

\clearpage

\appendix

\section{Quantum Numbers for Flavor Gauge Symmetry}
\label{app:Quan_nums}

The structure of Yukawa matrices in Eq.~(\ref{eq:quarkYukawa}) and Eq.~(\ref{eq:lepYukawa}) is ensured by the subtle symmetry introduced by an additional $U(1)$ transformation. Under this extended $U(1)$ group, the transformation behaviors of all relevant fields are given by
\begin{equation}
    \phi \to \phi' = e^{i\theta X_\phi} \phi,
\end{equation}
where the $U(1)$ charges required to maintain the desired Yukawa structures are chosen as follows,
\begin{align}
    & X_{Q_L} = \frac{1}{2} \begin{pmatrix}
        \mathcal{Q}_{u_R} + \mathcal{Q}_{d_R} & & \\
        & \mathcal{Q}_{u_R} + \mathcal{Q}_{d_R} & \\
        & & \mathcal{Q}_{t_R} + \mathcal{Q}_{d_R}
    \end{pmatrix},  
    X_{u_R} = \begin{pmatrix}
        \mathcal{Q}_{u_R} & & \\
        & \mathcal{Q}_{u_R} & \\
        & & \mathcal{Q}_{t_R}
    \end{pmatrix}, 
     \label{eq:charge} \\
    & X_{d_R} = \begin{pmatrix}
        \mathcal{Q}_{d_R} & & \\
        & \mathcal{Q}_{d_R} & \\
        & & \mathcal{Q}_{d_R}
    \end{pmatrix},\quad
    X_\Phi = \frac{1}{2} \begin{pmatrix}
        \mathcal{Q}_{u_R} - \mathcal{Q}_{d_R} & \\
        & \mathcal{Q}_{t_R} - \mathcal{Q}_{d_R}
    \end{pmatrix}, \quad X_S = \frac{1}{2} (X_{\Phi_1} - X_{\Phi_2}), \nonumber \\
    & X_{L_L} = \begin{pmatrix}
        \mathcal{Q}_{e_L} & & \\
        & \mathcal{Q}_{\mu_L} & \\
        & & \mathcal{Q}_{\tau_L}
    \end{pmatrix}, \quad
    X_{\ell_R} = \begin{pmatrix}
        \mathcal{Q}_{e_R} & & \\
        & \mathcal{Q}_{\mu_R} & \\
        & & \mathcal{Q}_{\tau_R}
    \end{pmatrix}, \quad X_{\nu_R} = 0. \nonumber
\end{align}
The $U(1)'$ charge matrices for different types of fermions are given by
\begin{align}
    & \mathcal{Q}_{uL} = \frac{1}{2} \begin{pmatrix}
        \mathcal{Q}_{u_R} + \mathcal{Q}_{d_R} & & \\
        & \mathcal{Q}_{u_R} + \mathcal{Q}_{d_R} & \\
        & & \mathcal{Q}_{t_R} + \mathcal{Q}_{d_R}
    \end{pmatrix} \label{eq:Z'charge}, \\
    & \mathcal{Q}_{dL} = \frac{1}{2} (\mathcal{Q}_{u_R} + \mathcal{Q}_{d_R}) \begin{pmatrix}
        1 & & \\
        & 1 & \\
        & & 1
    \end{pmatrix} + \frac{1}{2} (\mathcal{Q}_{t_R} - \mathcal{Q}_{u_R}) \begin{pmatrix}
        |c_1|^2 & c_1^* c_2 & c_1^* c_3 \\
        c_2^* c_1 & |c_2|^2 & c_2^* c_3 \\
        c_3^* c_1 & c_3^* c_2 & |c_3|^2
    \end{pmatrix}, 
    \mathcal{Q}_{uR} = \begin{pmatrix}
        \mathcal{Q}_{u_R} & & \\
        & \mathcal{Q}_{u_R} & \\
        & & \mathcal{Q}_{t_R}
    \end{pmatrix}, \nonumber \\
    & 
    \mathcal{Q}_{dR} = \mathcal{Q}_{d_R} \begin{pmatrix}
        1 & & \\
        & 1 & \\
        & & 1
    \end{pmatrix}, 
    \quad \mathcal{Q}_{\ell L} = \begin{pmatrix}
        \mathcal{Q}_{e_L} & & \\
        & \mathcal{Q}_{\mu_L} & \\
        & & \mathcal{Q}_{\tau_L}
    \end{pmatrix}, \quad
    \mathcal{Q}_{\ell R} = \begin{pmatrix}
        \mathcal{Q}_{e_R} & & \\
        & \mathcal{Q}_{\mu_R} & \\
        & & \mathcal{Q}_{\tau_R}
    \end{pmatrix}, \nonumber \\
    & \big(\mathcal{Q}_{\nu L}\big)_{ij} = \mathcal{Q}_{e_L} V^*_{ei} V_{ej} + \mathcal{Q}_{\mu_L} V^*_{\mu i} V_{\mu j} + \mathcal{Q}_{\tau_L} V^*_{\tau i} V_{\tau j}, \quad i,j = 1,2,3, \nonumber \\
    & \mathcal{Q}_{\nu R} = 0, \nonumber
\end{align}
where $c_i\equiv V_{ti}, i=d,s,b$ and
$d\equiv U_{u_R}|_{33}$. To obtain the above charges coupling to $ Z' $, we have made use of the structure of $ U_{uL} $ and $ U_{uR} $
\begin{equation}
	\begin{aligned}
		U_{uL}=\begin{pmatrix}
			\ast&\ast&0\\
			\ast&\ast&0\\
			0&0&1
		\end{pmatrix},~
		U_{uR}=\begin{pmatrix}
			\ast&\ast&0\\
			\ast&\ast&0\\
			0&0&1
		\end{pmatrix},
	\end{aligned}
\end{equation}
and 
\begin{equation}
	\begin{aligned}
		U_{\ell L}=U_{\ell R}={1}
	\end{aligned}
\end{equation}
which are required by the form of mass matrices. Apparently, FCNC occurs only in down-type quark sector with left handed chirality.

The exact charges of specific fermions are constrained by anomaly cancellation conditions, which yield the following relations,
\begin{align}
    & \mathcal{Q}_{u_R} = -\mathcal{Q}_{d_R} - \frac{1}{3} \mathcal{Q}_{\mu_R}, \quad
    \mathcal{Q}_{t_R} = -4 \mathcal{Q}_{d_R} + \frac{2}{3} \mathcal{Q}_{\mu_R}, \nonumber \\
    & \mathcal{Q}_{\tau_L} = \mathcal{Q}_{d_R} + \frac{1}{6} \mathcal{Q}_{\mu_R}, \quad
    \mathcal{Q}_{\mu_L} = -\mathcal{Q}_{d_R} + \frac{5}{6} \mathcal{Q}_{\mu_R}, \nonumber \\
    & \mathcal{Q}_{e_L} = \frac{9}{2} \mathcal{Q}_{d_R} - \mathcal{Q}_{\mu_R}, \quad
    \mathcal{Q}_{\tau_R} = 2 \mathcal{Q}_{d_R} + \frac{1}{3} \mathcal{Q}_{\mu_R}, \nonumber \\
    & \mathcal{Q}_{e_R} = 7 \mathcal{Q}_{d_R} - \frac{4}{3} \mathcal{Q}_{\mu_R}. \label{eq:charge2}
\end{align}
The above relations leave only two degrees of freedom, denoted as $\mathcal{Q}_{d_R}$ and $\mathcal{Q}_{\mu_R}$. The results presented here are consistent with those found in \cite{Celis:2015ara} by interchanging $e$ and $\tau$.

\section{Summary of rotation matrix}
\label{app:rotation_m}
The detailed rotation matrices between the mass basis and the interaction basis in Higgs potential are summarized as 
\begin{equation}
	\begin{aligned}
		U_{1}\equiv
		\begin{pmatrix}
			\cos\beta&\sin\beta\\
			-\sin\beta&\cos\beta
		\end{pmatrix},~
		U_{2}\equiv\begin{pmatrix}
			1&0\\
			0&U_\gamma
		\end{pmatrix}\begin{pmatrix}
			U_1&0\\
			0&1
		\end{pmatrix},~
		U_{3}\equiv\begin{pmatrix}
			-\cos\alpha&-\sin\alpha&0\\
			\sin\alpha&-\cos\alpha&0\\
			0&0&1
		\end{pmatrix},
	\end{aligned}
\end{equation}
where the $ 2\times2 $ matrix $ U_\gamma $ and $ \tan2\alpha $ have the explicit form 
\begin{equation}
	U_\gamma=\begin{pmatrix}
		\frac{2v_1v_2}{\sqrt{4v_1^2v_2^2+v^2v_S^2}}&\frac{-vv_S}{\sqrt{4v_1^2v_2^2+v^2v_S^2}}
		\\
		\frac{vv_S}{\sqrt{4v_1^2v_2^2+v^2v_S^2}}&\frac{2v_1v_2}{\sqrt{4v_1^2v_2^2+v^2v_S^2}}
	\end{pmatrix},~
	\tan2\alpha=\frac{2l}{m-n}.
\end{equation}
The rotation matrix defined in Eq.\eqref{eq:gauge_rotation} for explaining mass diagonalization procedure are
\begin{equation}
	\begin{aligned}
		U_{A \hat{Z}}=\begin{pmatrix}
			\cos\theta_{W}&-\sin\theta_{W}&0\\
			\sin\theta_{W}&\cos\theta_{W}&0\\
			0&0&1
		\end{pmatrix},\quad
		U_{\hat{Z} \hat{Z}'}=\begin{pmatrix}
			1&0&0\\
			0&\cos\theta_{2}^\prime&-\sin\theta_{2}^\prime\\
			0&\sin\theta_{2}^\prime&\cos\theta_{2}^\prime
		\end{pmatrix}\text{~and~}\tan2\theta_{2}^\prime= \frac{4|\Delta|}{4\delta-1}.
	\end{aligned}
\end{equation}

\section{Further Details on Phenomenology}
\label{app:pheo_detail}

The coefficients for the $ bsZ^{(\prime)} $ and $ Z^{(\prime)}\ell\ell $ vertices, extracted from Eq.~\eqref{eq:Zff} and utilized in Eq.~\eqref{eq:C9C10}, are summarized as follows,
\begin{equation}
    \begin{aligned}
        \mathcal{A}_{L}^{bsZ} &= g^\prime \sin\theta_{2}^\prime \mathcal{Q}_{dL}, \\
        \mathcal{A}_{L}^{bsZ'} &= g^\prime \cos\theta_{2}^\prime \mathcal{Q}_{dL}, \\
        \mathcal{B}^{Z\ell\ell} &= \frac{1}{2} \left[\frac{g_2 \cos\theta_{2}'}{\cos\theta_{W}} \left(-\frac{1}{2} + 2\sin^2\theta_{W}\right) + g'\sin\theta_{2}' (\mathcal{Q}_{\ell L} + \mathcal{Q}_{\ell R})\right], \\
        \mathcal{B}_{5}^{Z\ell\ell} &= \frac{1}{2} \left[\frac{g_2 \cos\theta_{2}'}{\cos\theta_{W}} \left(-\frac{1}{2}\right) + g'\sin\theta_{2}' (\mathcal{Q}_{\ell L} - \mathcal{Q}_{\ell R})\right], \\
        \mathcal{B}^{Z'\ell\ell} &= \frac{1}{2} \left[\frac{-g_2 \sin\theta_{2}'}{\cos\theta_{W}} \left(-\frac{1}{2} + 2\sin^2\theta_{W}\right) + g'\cos\theta_{2}' (\mathcal{Q}_{\ell L} + \mathcal{Q}_{\ell R})\right], \\
        \mathcal{B}^{Z'\ell\ell}_{5} &= \frac{1}{2} \left[\frac{-g_2 \sin\theta_{2}'}{\cos\theta_{W}} \left(-\frac{1}{2}\right) + g'\cos\theta_{2}' (\mathcal{Q}_{\ell L} - \mathcal{Q}_{\ell R})\right].
    \end{aligned}
\end{equation}
Here, it is noteworthy that the FCNC part $\mathcal{A}_{L}^{bsZ}$ from the SM $bsZ$ vertex is modified by the $U(1)'$ down-type charges due to mixing effects with the $bsZ'$ vertex.

Regarding the $bsZ/Z'$ vertex in $B_s$-$\bar{B}_s$ oscillations, in the SM only the single operator $\mathcal{O}_\text{VLL}$ is present at the loop level. In FG2HDM, from discussions involving $C_{9,10}$, it is inferred that no right-handed $\mathcal{O}_{\rm VRR}$ or $\mathcal{O}_{\rm LR}^1$ operators are generated due to the diagonal $\mathcal{Q}_{dR}$. These matrix elements in $\Delta M_s$ can typically be parameterized as bag parameters \cite{DiLuzio:2019jyq},
\begin{equation}
    \begin{aligned}
        \langle B_s^0 | \mathcal{O}_\text{VRR} | \bar{B}_s^0 \rangle = \langle B_s^0 | \mathcal{O}_\text{VLL} | \bar{B}_s^0 \rangle &= \frac{1}{3} m_{B_s} f_{B_s}^2 B_{1}, \\
        \langle B_s^0 | \mathcal{O}_{\text{LR1}} | \bar{B}_s^0 \rangle &= -\frac{1}{6} \mathcal{R}_{B_s} f_{B_s}^2 B_{5}, \\
        \langle B_s^0 | \mathcal{O}_{\text{LR2}} | \bar{B}_s^0 \rangle &= \frac{1}{4} \mathcal{R}_{B_s} f_{B_s}^2 B_{4}, \\
        \langle B_s^0 | \mathcal{O}_{\text{SRR1}} | \bar{B}_s^0 \rangle = \langle B_s^0 | \mathcal{O}_{\text{SLL1}} | \bar{B}_s^0 \rangle &= -\frac{5}{24} \mathcal{R}_{B_s} f_{B_s}^2 B_{2}, \\
        \langle B_s^0 | \mathcal{O}_{\text{SRR2}} | \bar{B}_s^0 \rangle = \langle B_s^0 | \mathcal{O}_{\text{SLL2}} | \bar{B}_s^0 \rangle &= -\frac{5}{6} \mathcal{R}_{B_s} f_{B_s}^2 B_{2} + \frac{1}{3} \mathcal{R}_{B_s} f_{B_s}^2 B_{3},
    \end{aligned}
\end{equation}
where $\mathcal{R}_{B_s} = \frac{m_{B_s}^3}{(\bar{m}_b + \bar{m}_s)^2}$ represents the generic factorized evolution of density-density operators.

\normalem

\bibliography{reference3}
\end{document}